\documentclass[12pt,preprint]{aastex}
\usepackage[dvips]{color}
\newcommand{\av}{A$_{\rm V}$}
\begin{document}
\title{Line Emission from Gas in Optically Thick Dust Disks around Young Stars}
\author{U.~Gorti\altaffilmark{1,2} }
\author{D.~Hollenbach\altaffilmark{2}} 
\altaffiltext{1}{University of California, Berkeley, CA}
\altaffiltext{2}{NASA Ames Research Center, Moffett Field, CA}   
\begin{abstract}
We present self-consistent models of gas in  optically-thick dusty disks and calculate its thermal,
density and chemical structure.  The models focus on an accurate treatment of the upper
layers where line emission originates, and at radii $\gtrsim 0.7$ AU. 
Although our models are applicable to stars of any mass, we present here 
only results around $\sim 1{\rm M}_{\odot}$ stars where we have varied dust properties, X-ray
luminosities and UV luminosities.
We separately treat gas and dust thermal balance, and
 calculate line luminosities at infrared and sub-millimeter wavelengths from
 all transitions  originating in the predominantly neutral gas that lies below the
 very tenuous and completely ionized surface of the disk. 
  We find that the [ArII] 7$\mu$m, [NeII] 12.8$\mu$m, [FeI] 24$\mu$m, 
[SI] 25$\mu$m, [FeII] 26$\mu$m, [SiII] 35 $\mu$m, [OI] 63$\mu$m 
and pure rotational lines of H$_2$ and CO can be quite strong and are good indicators of the
presence and distribution of gas in disks.  Water is an important coolant in the 
disk  and many water emission lines can be moderately strong. 
Current and future observational facilities such as the Spitzer 
Space Telescope, Herschel Observatory and SOFIA are capable of detecting gas 
emission from  young disks.
 We apply our models to the disk around the nearby young star, TW Hya, and
find good agreement between  our model calculations and observations. We also predict strong
emission lines from the TW Hya disk that are likely to be detected by future facilities.  
A comparison of CO
observations with our models suggests that
the gas disk around TW Hya may be  truncated to $\sim 120 $ AU,
compared to its dust disk of 174 AU. We speculate that  
photoevaporation due to the strong stellar  FUV field from TW Hya is responsible for the
gas disk truncation.
  \end{abstract}

\section{Introduction}	
Circumstellar disks  have been observed around stars with a wide range of  masses,
 from brown dwarfs to massive B stars  (e.g., Schreyer et al. 2006,
Luhman et al. 2007).  Disks facilitate accretion and angular momentum transport during the
star formation process and 
provide a mass reservoir of gas and dust for the formation of planetary systems. 
Planet formation in disks appears to be fairly common, with  observed extrasolar planets  now
totalling  over 250 objects (e.g., Udry et al. 2007).
The bulk of the initial gas and dust mass in disks around solar mass stars has a finite lifetime of order $\sim 1-10$ Myr (e.g., Haisch et al. 2001, Sicilia-Aguilar et al. 2006), many orders of magnitude smaller than  stellar lifetimes, and this sets 
an upper limit on the formation time of giant planets. 

Disk studies are largely based on measurements of the dust
emission. In primordial disks, although the dust mass is much smaller than the gas mass,  the dust 
is a strong emitter (see review by Watson et al. 2007). Disks 
 transition from a younger optically thick dust stage
characterised by strong infrared continuum excesses, to an optically thin
dust stage with weak infrared emission (e.g., Najita et al.  2007).
The census of optically thick young disks and disks
with scant infrared excesses like the weak-lined T Tauri stars (WTTS)  indicates that the
initial optically thick epoch lasts $\sim 3 $ Myr (e.g., Haisch et al. 2001), the transition
phase is rather abrupt ($\lesssim 0.1$Myr, see Hillenbrand 2007), and that the
WTTS phase also lasts for $\sim$ a few Myr.
 The missing small dust particles in WTTS and/or transition disks may
have coalesced to build planetary objects (e.g., Quillen et al 2004). 
However, observational studies of disks
infer initial masses ($\sim 0.1 {\rm M}_{\odot}$ or 100 M$_{\rm J}$)  much greater than
that of the mass of planets around stars ($\lesssim$ 10 M$_{\rm J}$),  suggesting that much of
the gas and dust is removed from the disk either by accretion onto the central star or 
by disk dispersal processes. 

The limited observations of gas in disks seem to broadly agree with the evolutionary sequence
traced  by  dust emission, but  dispersal timescales are less constrained at  $\lesssim$ 10 Myr (e.g., 
Zuckerman et al. 1995, Hollenbach et al. 2005,  Pascucci et al. 2006). 
Gas giant planets must necessarily form on these timescales before the gas disk disperses. 
 As gas dominates the mass of giant planets and  determines the dynamics of solid particles (until gas mass becomes smaller than particulate mass),  studies of the disk gas distribution and evolution are critical to understanding both planet formation and disk destruction processes. 

Gas disk observations are difficult  because of the inherently
weak line emission of the primary constituent, H$_2$. As a result, low abundance tracers are 
generally used to infer the presence of gas.  Gas emission lines have been observed
from numerous nearby disks, mainly millimeter and sub-millimeter wavelength
 rotational transitions of $^{12}$CO, $^{13}$CO,
HCO$^{+}$, DCO, HCN, CN, near-infrared vibrational lines of H$_2$ and CO, and more
recently mid-infrared H$_2$ pure rotational lines, 
as well as [FeI] and  [NeII] emission (e.g., review by Dutrey et al. 2007, Pascucci et al. 
2007, Lahuis et al. 2007).   The excitation of these
tracers  is sensitive to the density and temperature of the gas as well as
the chemical abundance of the emitting species. Therefore,  diagnosing gas disk parameters
from the measured line intensities  requires detailed  models of 
the thermal, density and chemical structure of gas in the disk. 

Detailed models of gas disks  are recently becoming available in varying degrees of sophistication
(e.g., Kamp \& Dullemond 2004, Gorti \& Hollenbach
2004, hereafter GH04, Jonkheid et al. 2004, Glassgold et al. 2004, 2007,
 Nomura \& Millar 2005, Aikawa \& Nomura 2006, Nomura et al. 2007).
Predicted line emission from  these models can be compared with 
observations to infer the physical conditions in disks.  Many existing
gas disk models solve for the chemistry and gas temperature,
but assume a disk vertical density structure
that is derived from modeling the dust radiative transfer
 (e.g.,  Glassgold et al. 2004, 2007,  Semenov et al. 2006, Meijerink et al. 2007).   Kamp \& Dullemond (2004)
calculated the gas density and temperature by adopting the disk structure
determined by the dust models of Dullemond et al.  (2002), whereas
Jonkheid et al.  (2004) and Glassgold et al. (2004, 2007)
use the dust disk model of D' Alessio et al. (1998).  
Because the gas temperature may differ from the dust temperature and the gas/dust
ratio may change with vertical height, these calculations may be inconsistent with the
condition of vertical pressure equilibrium. 
These inconsistencies are especially severe in the upper layers of disks where gas and dust temperatures can deviate significantly. Moreover, most observed gas emission originates
from these layers making any interpretation of data contingent on obtaining the correct gas
temperatures and densities at these heights.
GH04 and Nomura \& Millar (2005)  address this issue and self-consistently calculate the
gas density and temperature as a function of disk radius and vertical height.
  In these models, the dust temperature
is separately determined and the gas temperature obtained from 
a detailed thermal balance and chemistry calculation, by considering
various heating and cooling sources. GH04 only consider optically
thin dust disks, but solve for the gas disk in detail with heating by
dust collisions, X-rays, grain photoelectric heating due to far ultraviolet
(FUV) radiation incident on very small grains, cosmic rays and 
exothermic photoreactions, and cooling by many atomic, ionic and
molecular transitions. Nomura \& Millar (2005) consider  younger disks
with optically thick dust, but their gas model is more simplistic. They
consider heating by dust collisions, cosmic rays, and grain photoelectric heating,
 and cooling is only by OI, CII and CO lines. 
Aikawa \& Nomura (2006) recently extend these models to compute detailed disk chemistry, but use a disk structure (density and temperature) derived from the earlier simpler models, and Nomura et al. (2007) include the effects of X-ray heating. 

Our earlier models (GH04) of gas emission from evolved, optically thin dust disks combined with
recent   {\it Spitzer} observations of disks around stars of various ages
(FEPS Legacy Science Program), found that disks with
optically thin dust also have very little gas (Pascucci et al. 2006).
We could set  stringent limits on the gas mass using our models in conjunction with this
data and estimated upper limits on gas disk lifetimes to be $5-30$ Myrs. 
This result suggests that as  inner ($\sim 1-20$ AU)  disks transition from optically thick
to optically thin,  they also lose much of their mass in gas.  
The same result was  independently determined  by a recent study of a large sample of disks in Chameleon, where disk gas accretion rates were found to be correlated with the presence of an inner dust disk, suggesting that dust and gas are removed simultaneously from the inner disk (Damjanov et al. 2007).
 Gas dispersal therefore presumably begins at earlier stages of evolution when the disks are still optically thick in dust. 

In this paper, we extend our earlier detailed gas disk models (GH04)
 to consider younger disks with optically thick dust.   We are primarily concerned with
 calculating line emission from young disks that can be used as
gas diagnostics to trace its distribution and structure at different evolutionary stages
during its optically thick dust epoch.  We also propose to use these disk models to
compute the mass loss rate due to photoevaporation from stellar EUV, FUV and
X-ray photons in an accompanying paper.  Our models are constructed with the
necessary physics to determine the disk thermal and chemical structure in the
surface layers of the disk where line emission arises and photoevaporative flows
originate. 

 The structure of the paper is as follows. We describe the details of our disk models in \S 2. We  discuss the resulting disk structure  and line emission in  \S 3.  In \S 4, we apply our models to observational data from the well-studied disk of TW Hya and discuss  the results. We summarize and 
conclude in \S 5.

\section{Model description, inputs and assumptions}
\subsection{Main model features}
The disk models presented here are an extension of GH04.  Our earlier models were
restricted  to cases where the dust was optically thin to stellar radiation, keeping our dust radiative transfer simple. However, the gas transitions in these models could be optically thick or thin depending on gas column density, and gas radiative transfer was calculated using an escape probability 
formalism. In this paper, we generalize our models to consider optically thick dust and therefore
 allow  modeling of younger disks which are opaque in the midplane regions
 to stellar visible photons.  In most of our models, disks are also opaque to
 their own infrared photons. Other 
improvements are the addition of  new coolants such as fine structure lines of 
Ne and Ar ions  and vibrational lines of CO,  heating by and chemistry of Polycyclic Aromatic Hydrocarbons (PAHs), a mixture of dust chemical compositions, and a model for the EUV heated surface layer. We treat in a separate paper the emission of fine structure lines such
as [NeII]$12.8\mu$m from the EUV-heated and completely ionized
surface layer, and here only use the model
to determine the vertical position of the ionization front that separates this layer from the 
predominantly neutral disk underneath it (Hollenbach et al. 1994).  In this paper, we
do, however,  treat the
ionized fine structure lines such as [NeII]$12.8\mu$m
produced in the ``neutral disk'' where X-rays partially ionize
and heat the gas. 

We briefly summarize the disk model here and in the remainder of this section provide more details for the interested reader.  Our disk models  solve for chemistry and thermal balance, and impose vertical hydrostatic equilibrium to separately
calculate the density and gas and dust temperatures as a function of spatial location in the disk.
We consider heating of the gas due to X-rays, grain photoelectric heating by PAHs and 
small grains, cosmic rays, exothermic chemical reactions, formation heating of H$_2$, collisional de-excitation of vibrationally excited (by FUV) H$_2$, photodissociation and photoionization of
molecules and atoms,
 and collisions with warmer dust grains.  Cooling of gas is by line emission from atoms, ions
and molecules, and by collisions with cooler dust grains. Our chemical network is moderate and 
focused towards including species that
are dominant coolants in the disk (enabling an accurate determination of the gas temperature).
We treat gas and dust independently, and allow for different spatial distributions (i.e., a 
position-dependent gas/dust mass ratio),
although we generally consider them well-mixed and the ratio constant throughout the disk. We consider a mixture
of chemical compositions for the dust, and a range of grain size distributions. Dust radiative
transfer for optically  thick dust disks is simplified, so as to keep the numerical disk model computations 
tractable. We use the two-layer model for dust with some
modifications as first described by Chiang \& Goldreich (1997).  We include the effects of  background infrared radiation due to dust on line transitions
of the gas species (see Hollenbach et al. 1991). We emphasize that our models are hydrostatic, and assume steady-state chemistry and thermal balance. We also neglect freezing of  gas species on to cold dust grains. This process is likely to be important only in the dense midplane  of the outer disk (e.g., Aikawa et al. 1999, 2002),  a region unimportant for many of the considerations of this paper.  
Accretion heating can be important in the dense midplane, especially  of the inner disk (e.g., D'Alessio et al. 1998). We ignore this term, however, because
we are interested in evolutionary stages of the disk when the accretion rates are low
 and  because gas emission lines mostly originate from the upper layers of the disk.  In the modeling procedure, the input parameters include stellar characteristics,
the initial gas phase abundances of the 10 elements, the surface density distribution, radial extent and dust properties. We then calculate the gas temperature, density and chemical structure as a function of spatial location $(r,z)$ throughout the disk. 

\subsection{Stellar Characteristics}
The properties of the central star are basic inputs to our disk models.  When modeling observations of specific star-disk systems such as TW Hya in \S 4,  stellar properties are taken from the available literature.  In \S 3 of this
paper, we aim to be more general and use typical stellar characteristics to construct standard or fiducial disk models that depict our calculations of the structure and line emission.

The models presented in this paper assume {\em steady-state} thermal and chemical balance at a time when the disk surface density distribution follows a prescribed power law
with radius. They are not time-dependent calculations. 
However, stellar properties evolve with 
time, the disk mass and surface density distribution evolve, and the accretion rate onto the central star decreases (Hartmann et al. 1998) with time.  Younger stars have stronger radiation fields 
which would result in greater heating and enhanced emission line strengths.  For our fiducial disk model, 
we use  a {\em median} set of stellar characteristics during this evolution,  typical of a 1-2 Myr old system.
 We use the pre-main-sequence evolutionary tracks of Siess et al. (2000) to establish the basic properties of the star, such as its mass, radius, bolometric  luminosity and effective temperature.  
 In this paper, we focus on disks around  1M$_{\odot}$ stars. 

The radiation field created by the central star greatly affects  the disk structure.  Disk models to date have primarily focused on the dust component which absorbs the optical photons  from the star, and  optical spectra of young stars are very well determined.   Gas is then assumed to be collisionally coupled with the dust and follow the same temperature distribution.  However, in our models, we  {\em independently} calculate the gas temperature. Apart from indirect heating by stellar optical photons via  collisions with the dust warmed by these photons, gas can also be heated by the ultraviolet and X-ray flux from the star. These photons also influence gas chemistry which can affect disk structure indirectly by influencing the cooling which sets the gas temperature structure
(see GH04 for details on these processes).  Hence we need to characterise the stellar spectrum at these shorter wavelengths, which are less well determined than the optical spectra. 

\paragraph{X-rays} Recent observational studies using ROSAT and Chandra have significantly furthered our understanding of the nature of the X-ray emission and pre-main-sequence magnetic activity of young stars (e.g., Feigelson \& Montmerle 1999, the COUP Project, Feigelson et al. 2005, Preibisch et al. 2005). The X-ray luminosity of low mass young stars is believed to mainly originate in the chromosphere and observed to be correlated with the bolometric luminosity, $\log L_X/L_{bol} \sim -3.6$, although with a large scatter (Preibisch et al. 2005).  For more massive stars, (M$_*  \gtrsim 3 {\rm M}_{\odot}$) the X-ray luminosity decreases and  $\log  L_X/L_{bol} \sim -6$ (Preibisch et al. 2005, Steltzer et al. 2005). We therefore scale the total X-ray luminosity of a star with its bolometric luminosity according to these empirical relations.  X-rays originating from accretion processes peak at lower energies, and new Chandra and XMM-Newton X-ray spectroscopy of  a few nearby, young 
accreting stars reveal  both a low and high energy component (Telleschi et al. 2006).  Since the X-ray absorption cross-section increases 
with decreasing photon energy (Wilms  et al.  2000), a softer X-ray spectrum would heat the lower density gas at the disk surface, whereas harder X-ray photons will penetrate deeper into the disk
to heat higher density gas. We therefore extend our X-ray spectrum from 0.1 to 10 keV. Our adopted X-ray spectrum (G\"{u}del et al. 2007, Feigelson \& Montmerle 1999, also see GH04) peaks at 2 keV and is approximated by a simple power-law with $dL_X/dE  \sim E$ for $0.1 < E < 2$ keV and $dL_X/dE  \sim E^{-2}$ for $ 2<E<10$ keV.

\paragraph{UV} 
FUV (6eV$< h \nu<$13.6eV) spectra of young stars are more difficult to determine observationally.  The spectra are also highly variable, heavily extincted  and of somewhat uncertain origin 
 (e.g., IUE, Valenti et al. 2003; FUSE spectra, e.g., Bergin et al. 2003, Herzceg et al. 2004). The FUV flux appears correlated with the accretion rate, suggesting that accretion hotspots are the main cause of the UV excess, whereas significant FUV fluxes from low accretors also indicate a chromospheric component (Calvet \& Gullbring 1998,  Valenti et al. 2003). For the purpose of this study, we follow the results of the various studies by Gullbring et al. (1998, 2000)  and assume that the FUV originates from accretion hotspots on the surface of the star. We use the empirical results of Muzzerolle et al. (2003) which suggest  that the mass accretion rate $\dot{M}_{acc} $  for a given stellar mass scales with $M_*^2$.  For a $1  {\rm M}_{\odot}$ star of age $1\sim2$ Myr, we take a typical 
$ \dot{M}_{acc}  \sim 3 \times 10^{-8} {\rm M}_{\odot}$ yr$^{-1}$. 
 We then calculate the accretion luminosity ($L_{acc} = G M_* \dot{M}_{acc} / 2 R_* $).  The FUV spectrum is assumed to be a blackbody at a ``hotspot'' temperature of 9000 K (e.g., Gullbring et al. 1998, 2000).  We independently confirmed the validity of the above approach by estimating the median FUV flux from stars with good quality IUE spectra (Valenti et al. 2003, IUE spectral atlas). In both cases, the median $L_{FUV}/L_{bol}$ ratio for solar-mass stars was found to $\sim 10^{-2}$ to $10^{-3}$. We, however, adopt the semi-empirical approach of an accretion-rate determined FUV flux and spectrum, because of the uncertainties (extinction, distances, stellar mass, age)  in converting the observed IUE spectra to a mass-dependent FUV flux. 
  
The EUV luminosity of young solar mass stars is very poorly constrained by observations,
but believed to be generated by the active chromospheres from these stars
(e.g., Bouret \& Catala 1998, Alexander et al. 2005). We
therefore assume an EUV luminosity similar in magnitude to that of the chromospheric FUV and
X-ray luminosities,  $\approx 10^{-3} $ L$_{bol}$. The photon luminosity $\phi_{\rm EUV}$ 
is thus $ \sim 4 \times 10^{41}$ s$^{-1}$ for a  1M$_{\odot}$ star, which then determines the location of the ionization front
at the disk surface (e.g., Hollenbach et al. 1994). 


Stellar parameters of our fiducial disk model are listed in Table~\ref{starpar}. 

\subsection{Dust properties}

Thermal energy exchange by collisions of gas with dust is one of the key processes in disks,
 which even at their surfaces (${\rm A}_{\rm V} \sim 1$ to the central star) have high gas densities
 ($n \sim 10^7$ cm$^{-3}$) compared to interstellar clouds.
 Dust grains at the disk surface absorb stellar optical 
radiation, warming this surface layer.  This  warm thin layer with very little mass then 
heats the disk interior through re-radiation (Chiang \& Goldreich 1997, D'Alessio et al. 1998, Dullemond, Dominik \& Natta 2001). Collisions with dust can either heat or cool the gas depending on whether the dust grains are hotter than or cooler than the gas.  If they are well-coupled, gas and dust temperatures are nearly identical. However, the collisional coupling may be
reduced by dust settling to the midplane, a process that can significantly decrease the local ratio of dust to gas mass ($\ll 0.01-$ the interstellar and disk primordial  value) and the dust cross sectional area per H nucleus, $\sigma_{\rm H}$,  near the surface of the disk. In addition, grains may have coagulated, a process that also decreases $\sigma_{\rm H}$. 
The decrease in collisional coupling enhances the difference between the gas and dust temperatures (GH04, Kamp \& Dullemond  2004, Jonkheid et al. 2004) because of other possible 
processes that can affect the thermal balance of the gas. This is  especially true at the disk surface where  densities and opacities are lower.
Gas heating/cooling and disk structure thus depend on dust properties and are affected by both grain growth and settling. 

We consider dust grains to have a range of sizes,  distributed according to the MRN power-law characteristic of dust
in the interstellar medium ($n(a) \propto a^{-3.5}, a _{min} < a < a_{max}, $ where $a$ is the grain radius). We  assume spherical grains and an interstellar dust composition. Grain growth up to 
at least millimeters in size is inferred to
occur in young circumstellar disks (e.g., van Boekel et al. 2005,
 Shuping et al. 2006, Muzerolle et al. 2006, Kessler-Silaci et al. 2006).
Dust settling is also inferred with settling timescales in disks  calculated to be rather short  ($\sim 10^4$ years at 1 AU, Dullemond \& Dominik 2004), but observations indicate the presence of small dust even in the surfaces of evolved disks (Hernandez et al. 2005, Furlan et al. 2006), suggesting the operation of other mechanisms that cause shattering of grains and the mixing of grains from the midplane to the surface (e.g., Dullemond \& Dominik 2005).  We keep the minimum and maximum grain sizes constant in this work at 50\AA \  and  20$\mu$m respectively,  which lowers the opacity 
($\sigma_{\rm H}$) by a factor of $\sim 10$ from interstellar values (where $\sigma_{\rm H}
 \sim 2 \times 10^{-21} {\rm cm}^{2}$). 
Grain growth in the disk will lead to an increase in $a_{max}$ in general. One would also expect that since small grains coagulate to form larger objects, $a_{min}$ might also increase. However, as pointed out earlier, observational studies infer the presence of small dust grains (and PAHs) in evolved disks(e.g., Hernandez et al. 2005, Furlan et al. 2006, Geers et al. 2006)  presumably formed by a collisional cascade process which replenishes the small dust population in the disk (e.g., Weidenschilling \& Cuzzi 1993).  We therefore fix $a_{min}$ in our models. The important physical parameter in the models is the dust area per H atom, $\sigma_{\rm H}$, which can be reduced either by increasing the dust grain sizes {\em or} by reducing the  dust/gas mass ratio ($\eta$) in the disk.  In our modeling of dust properties,  we have already assumed some grain growth by increasing $a_{max}$ from the interstellar value of $\sim 0.2\mu$m to a value of $20\mu$m. However, we model further possible variations in $\sigma_{\rm H}$ due to either further grain growth and/or settling by varying the dust/gas ratio, $\eta$.  This allows us to include dust settling by setting $\eta$ to be a function of position ($r,z$) in the disk. However, in this paper, we keep the ratio $\eta$ a constant for any given model.  To summarize,  we model  reductions in opacity due to grain growth or settling processes first by increasing $a_{max}$ to $20\mu$m and then subsequently by simply  reducing the dust/gas mass ratio in the disk from its fiducial (and interstellar) value of $0.01$.
 We consider one additional  case with $\sigma_{\rm H}  \sim 2 \times 10^{-24} {\rm cm}^{2}$, or a reduction factor of 100 from our fiducial case.  Note that although modeling grain growth/settling by reducing $\eta$ does not affect gas heating, temperature or spectra, which depend only on the gas-grain coupling ($\sigma_{\rm H}$), the dust continuum emission would be affected by this procedure. When modeling specific star-disk systems, such as TW Hya (\S 4), we use a dust grain size distribution constrained by the continuum dust emission.

Polycyclic Aromatic Hydrocarbons or PAHs are important gas heating agents in the
FUV-irradiated ISM. They heat the gas via FUV photoelectric ejection of energetic
electrons into the gas.  PAHs are abundant in the interstellar 
medium and hence are likely to be present in primordial disks in the very earliest stages of disk 
evolution. The presence of PAHs in disk surfaces is indicated by their observation in disks around both low-mass T Tauri stars and intermediate-mass HAeBes (Habart et al. 2004, Acke and van Ancker 2004, van Boekel et al. 2004, Geers et al. 2006).   It is unclear how the PAH abundance is affected by the dust evolution process and in view of these recent studies we assume that disks do contain PAHs, but  in reduced abundances. We calculate the reduction in $\sigma_{\rm H}$ for our assumed dust grain size distribution as compared to the ISM, and scale
the PAH abundance down from that in the ISM ($8.4\times10^{-7}$, Li \& Draine 2001)
 with the same factor. 
 PAHs can dominate gas heating, especially in the 
presence of strong UV fields that are likely to be present during early disk evolution 
epochs. We also consider collisional heating of gas by warm PAH molecules. 
We follow Li \& Draine (2001) and Draine \& Li (2001) for calculating the PAH temperature and
absorption coefficients and Weingartner \& Draine (2001) for the heating due to photoelectric
effect by small grains and PAHs. 

\subsection{Disk properties}
For our model calculations, the disk mass, gas/dust ratio, radial extent and surface density distribution are input parameters.  Model disk masses, motivated by the above studies,  are assumed to  scale with the mass of the central star and we adopt $M_{disk}/M_* \sim 0.03$ (e.g.,
Beckwith et al. 1990, Natta et al. 2001, Andrews \& Willams 2005).
  We also assume a simplistic surface density distribution, given by a power law, $\Sigma(r) \propto r^{-p}$, with $p=1$, as is likely for a disk with a constant viscosity parameter $\alpha$ in a steady state of viscous accretion (e.g, Hartmann et al. 1998). The dust inner radius is fixed at 0.5 AU.
  However, we do not accurately treat the inner rim in our simple dust model, and
  hence only report emission from beyond $r = 0.7$ AU where we believe our results
  to be accurate.  We also calculate an additional disk model with an inner radius of
  0.1 AU to test the sensitivity of line  emission  to this input parameter.
  We adopt a disk size of $200$ AU, as is typical for observed disks (e.g., Andrews \&
  Williams 2007). 
  
\subsection{Thermal balance, chemistry and gas temperature}
The gas heating processes included are collisions with warmer dust grains, grain photoelectric heating by PAHs and small dust grains, X-rays, cosmic rays, photochemical reactions, FUV pumping of H$_2$
and exothermic reactions. We include a modest chemical network (84 species, $\sim 600$ reactions,
see GH04) that involves the dominant chemical species of H, He, C, O, Ne, Mg, Fe, Si, Ar and S and solve for the gas chemistry (see Table~\ref{abun} for adopted gas phase abundances of these elements).  We have added Ne and Ar (since GH04), as they can dominate the cooling in the upper, X-ray heated layers of the disk (e.g., Glassgold et al. 2007).  We adopt similar procedures as Glassgold et al. for ionization of  Ne (and Ar), where X-rays are assumed to doubly ionize the species, which then can recombine with electrons (or undergo charge exchange reactions with H) to form singly ionized species. We ignore ionizations to higher levels as the electron recombination rates to lower ionized states are typically high 
and most of the X-ray ionizations of the predominantly neutral species lead to
the doubly ionized state (e.g., Maloney et al. 1996, Glassgold et al. 2007). H$_2$ forms on grain
surfaces, by reaction of H with H$^-$,  and  by three-body reactions. The latter are 
especially important for disks with a low $\sigma_{\rm H}$. The chemistry in our models closely follows that of photodissociation regions or PDRs (Tielens \& Hollenbach 1985, Kaufman et al. 1999), with some modifications for geometry and the high gas densities in disks (see GH04). Photodissociation of H$_2$ and CO by FUV and X-ray photons and their self-shielding is calculated as described in Tielens \& Hollenbach (1985) and Maloney et al. (1996) and described in more detail in GH04.

 Cooling processes include collisions with colder dust particles and line transitions of ions, atoms and molecules. 
Radiative gas cooling is usually dominated by transitions in the infrared of various chemical 
species, and we have used the approach outlined in Hollenbach et al. (1991) to estimate the background infrared radiation field (from dust re-radiation) and to calculate its effect on the level populations of the coolants.  We refer the reader to GH04 for other details of the heating, cooling and chemistry. 

 Gas line emission and photoevaporative flows typically originate in the upper layers of the disk, where the visual extinction to the star A$_{\rm V} \lesssim 1$.  The physics and chemistry included in our models therefore mainly focus on calculating the disk structure in the upper regions accurately. We do not consider some processes likely to be important in the very dense midplane regions of the disk, such as accretion heating (likely to be important in the midplane at $r\lesssim$ few AU) for a standard MRI driven  $\alpha$-disk model and freezing of chemical species on cold dust grains (important in the midplane regions of the outer $r\gtrsim 5$ AU disk, e.g., Aikawa et al. 1999, Markwick et al. 2002,
 Davis 2007). 

\subsection{Disk structure}
We construct 1+1D models of the gas disk with the surface density distribution and 
radial extent of the disk specified. For simplicity, the gas and 
dust are assumed to be well-mixed, i.e., in the work presented in this paper, we 
assume a uniform gas-to-dust ratio throughout the disk and ignore the fact that settling may result in an increase of the gas/dust ratio with height. The gas pressure
is set by vertical equilibrium between thermal pressure and gravity in the disk. 
Density and temperature are determined to be consistent with the prescribed
surface density distribution, pressure, as well as chemical and thermal balance. 
Appendix A provides some details of the numerical scheme for obtaining a disk solution. The numerical models for the gas 
thermal, density and chemical structure are computationally intensive, and we therefore
simplify our dust radiative transfer to a two-layer model (see Chiang \& Goldreich 1997, D'Alessio et al. 1998, Dullemond, Dominik \& Natta 2001, Rafikov \& De Colle 2006).  This simple 
approximation is fairly accurate in reproducing the  disk structure as 
compared to full 2D calculations (e.g., Dullemond 2002).  We use the correction factors
prescribed by Dullemond, Dominik \& Natta (2001) to account for the possibility of
the disk becoming optically thin in the infrared, smoothly transit dust temperature from interior to the surface, and use the prescription by Rafikov \& De Colle (2006) for determining the grazing angle to the disk surface.  
The vertical extent of the model disk is equal to the local radius, approximately 6-7 scale heights, where the visual extinction to the star is almost negligible.

The stellar EUV field will create an ionized HII region above the optical, FUV and X-ray 
 heated neutral layer  and we do not explicitly solve for the thermal structure of this ionized layer. Instead, we use the analytical formulae of Hollenbach et al. (1994) 
for the density of the ionized gas at the ionization front (IF) and assume a fixed temperature of $10^4$ K for the ionized gas above this front. 
 We ensure that
our FUV and X-ray heating calculations extend to a height where the pressure is equal to
that at the base of the ionized EUV-heated layer (i.e., the IF).  The gas density in the EUV-heated layer is again determined from vertical pressure equilibrium, and ionization dominates gas chemistry in this layer. 

We take the analytic equation for the electron density $n_b(r)$ at the base of this layer
(i.e., at the IF, below which lies the predominantly neutral disk) from
Hollenbach et al. (1994)
\begin{equation}
n_b(r) =  n_{b0} \left( r/r_g \right)^{-p}
\label{dbase}
\end{equation}
where 
\begin{equation}
n_{b0} = C_1 \left( { {3 \phi_{EUV}} \over{4 \pi \alpha_r r_g^3}}\right)^{1/2} = 
2.5 \times 10^5 \phi_{42}^{1/2} \left( { {{\rm M}_*} \over{1 {\rm M}_{\odot}}}
\right)^{-3/2} {\rm cm}^{-3}\end{equation}
\label{dbase0}
\begin{equation}
r_g = 7.3 \left( { {{\rm M}_*} \over{1 {\rm M}_{\odot}}}\right) {\rm AU}
\end{equation}
and where $C_1=0.293$ is a numerical correction factor,
$\phi_{EUV} = 10^{42} \phi_{42}$ s$^{-1}$ is the EUV luminosity of the star,
$\alpha_r=2.57\times 10^{-13} {\rm cm}^{3} {\rm s}^{-1}$ is the electron recombination
coefficient for H$^+$ at $10^4$ K, and $r_g$ is the characteristic gravitational radius where the sound
speed in the HII region equals the escape speed from the disk/star system. The
exponent $p$ in Eq.~\ref{dbase} is $3/2$ for $r\leq r_g$ and $5/2$ for $r\geq r_g$. 

The thermal pressure at the base of the EUV region (in units of $n$T) is then given by 
$P_b(r) \simeq 2 \times 10^4 n_b(r) {\rm cm}^{-3}$ K, or 
\begin{equation}
P_b(r) = 5 \times 10^9 \phi_{42}^{1/2} \left( { {{\rm M}_*} \over{1 {\rm M}_{\odot}}}\right)^{-3/2}
\left( {{r}\over{r_g}}\right)^{-p} {\rm cm}^{-3} {\rm K} 
\end{equation}
where  $p$  is $3/2$ for $r\leq r_g$ and $5/2$ for $r\geq r_g$.

In our disk model, we calculate the neutral gas density and temperature vertically, 
and insert the IF  when the pressure becomes equal to $P_b(r)$.
At that height,  we set the gas temperature to $10^4$K and the H nucleus density to
$n_b(r)$.  Because of the sharp rise in temperature across the IF, $n$ and \av \ 
drop abruptly across the IF. 
Above the base of this IF, the density drops as
\begin{equation}
n(r) = n_b(r) e^{-z^2/2 H(r)^2}
\end{equation}
with the scale height $H(r)$ given by $H(r)=r_g (r/r_g)^{3/2}$ for $ r\leq r_g$ and
$H(r)=r$ for $r\geq r_g$ (see Hollenbach et al. 1994).
The vertical hydrogen nucleus column in the EUV layer is then
\begin{equation}
N_{EUV} \simeq H(r) n_b(r) = 3 \times 10^{19} \phi_{42}^{1/2} 
\left( { {{\rm M}_*} \over{1 {\rm M}_{\odot}}}\right)^{-1/2} {\rm cm}^{-2} {\rm \ for\ }  r\leq r_g.
\end{equation}
Note that this column is independent of $r$ for $r< r_g$, and, assuming $\sigma_{\rm H}
=2 \times 10^{-22} {\rm cm}^2/{\rm H}$, $\phi_{42}=1$, and M$_*=1{\rm M}_{\odot}$, 
corresponds to a vertical \av\ from the base to infinity of only $\sim 1.5 \times 10^{-3}$. The
column from the base to the star is given by
\begin{equation}
N_{b*} \simeq r n_b(r) = 3 \times 10^{19} \phi_{42}^{1/2} 
\left( { {{\rm M}_*} \over{1 {\rm M}_{\odot}}}\right)^{-1/2} \left({ {r}\over {r_g}}\right)^{-1/2}
 {\rm cm}^{-2}.
 \end{equation}
 Therefore, at $r=r_g$, the \av \ from the base of the IF (or the base of
 the EUV layer) to the star is also $\simeq  1.5 \times 10^{-3}$ for our standard dust
 model and $\phi_{42}=1$, and M$_*=1{\rm M}_{\odot}$.

The different input parameters and assumptions of our fiducial disk model are summarized for convenience in Table~\ref{fidpar}.

\section{Results and discussion }

\subsection{Disk structure and chemistry}
\begin{figure}
\plotone{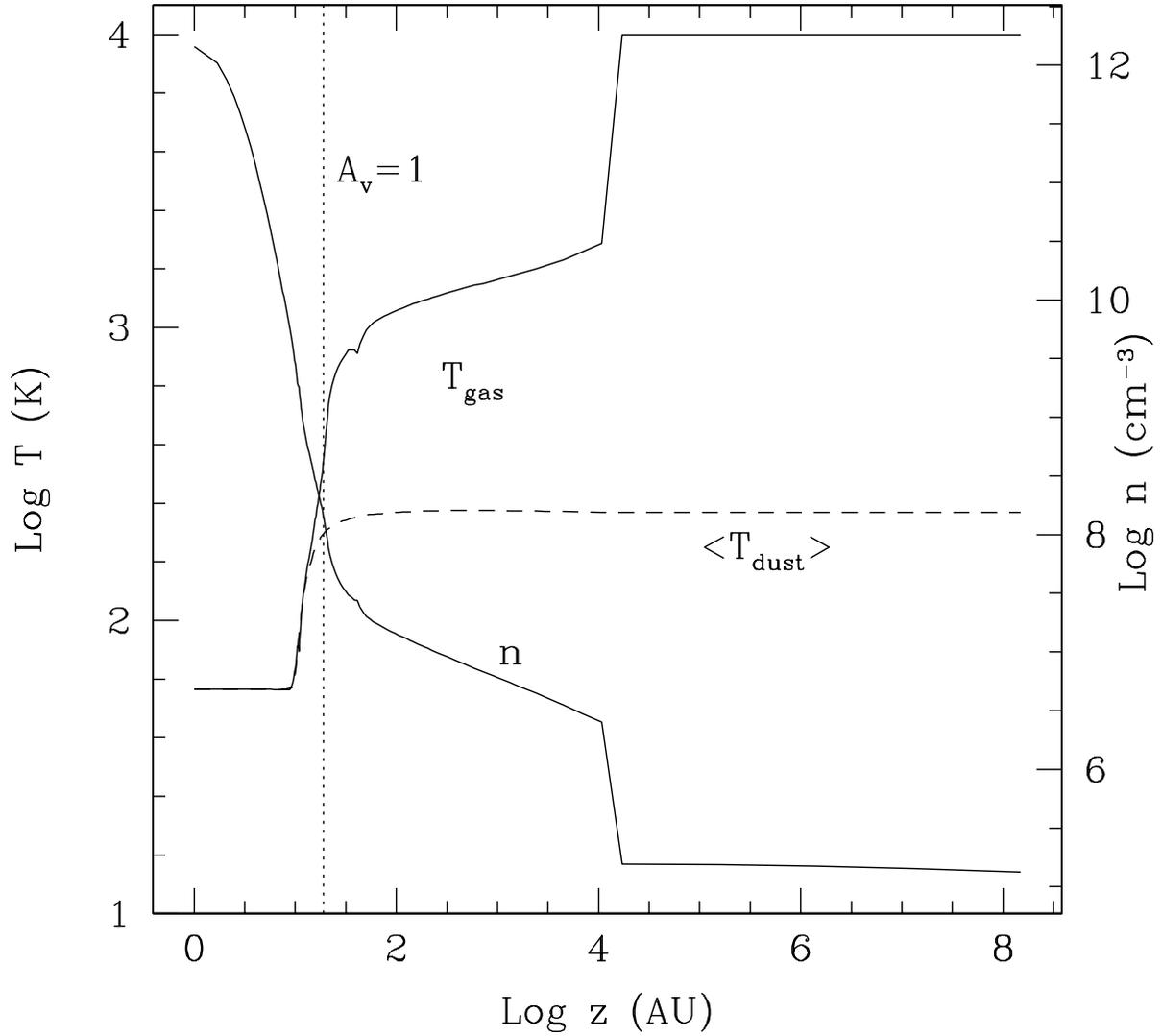}
\caption{Vertical temperature and density profile at 8 AU. The gas temperature
and density are shown by solid lines, and the mean dust temperature is depicted
by the dashed line. The dotted vertical line shows the location of the height where the
visual extinction \av \ to the star is unity. }
\label{fidtemp}
\end{figure}

\begin{figure}
\plotone{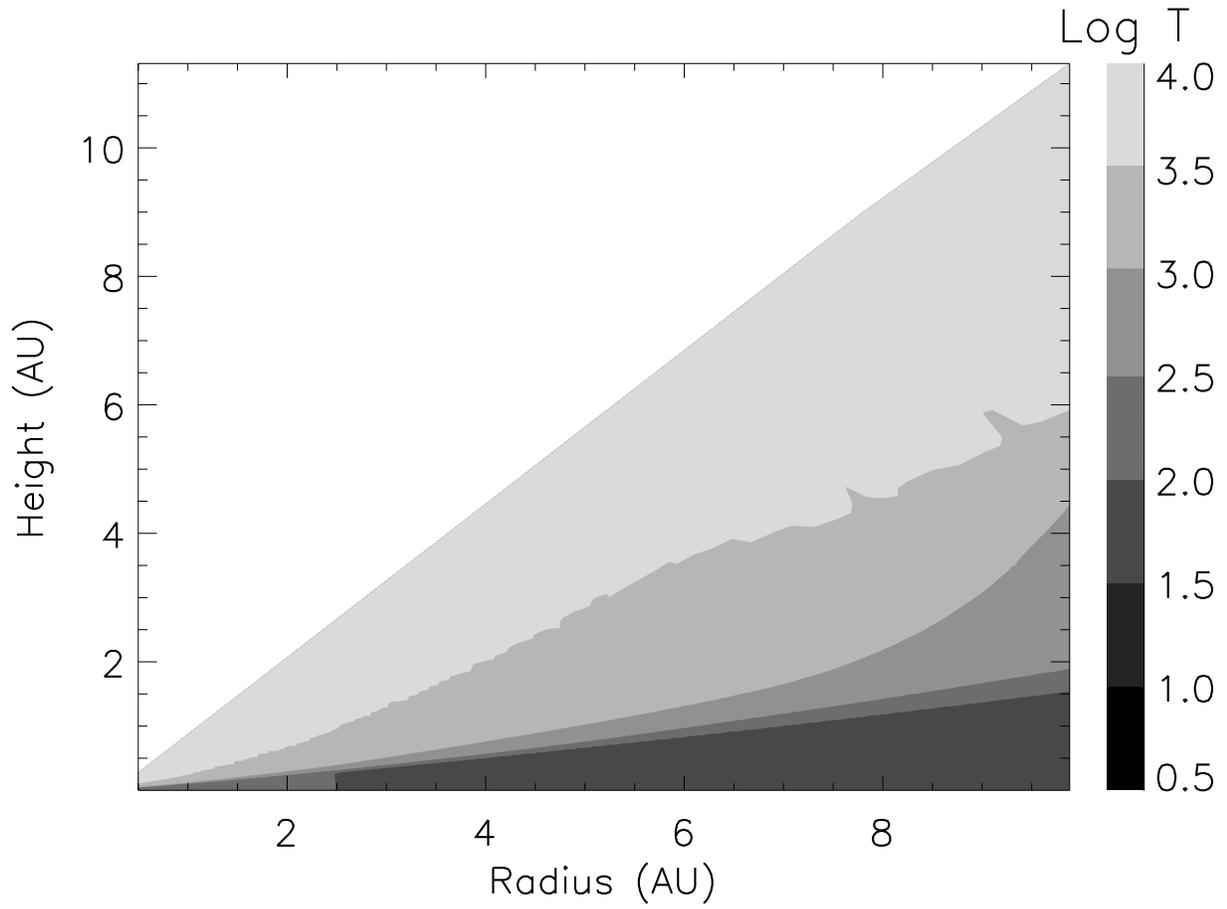}
\caption{Gas temperature contour plot of the inner 10 AU region of the
fiducial disk (Model A).  }
\label{tempcon20}
\end{figure}

\begin{figure}
\plotone{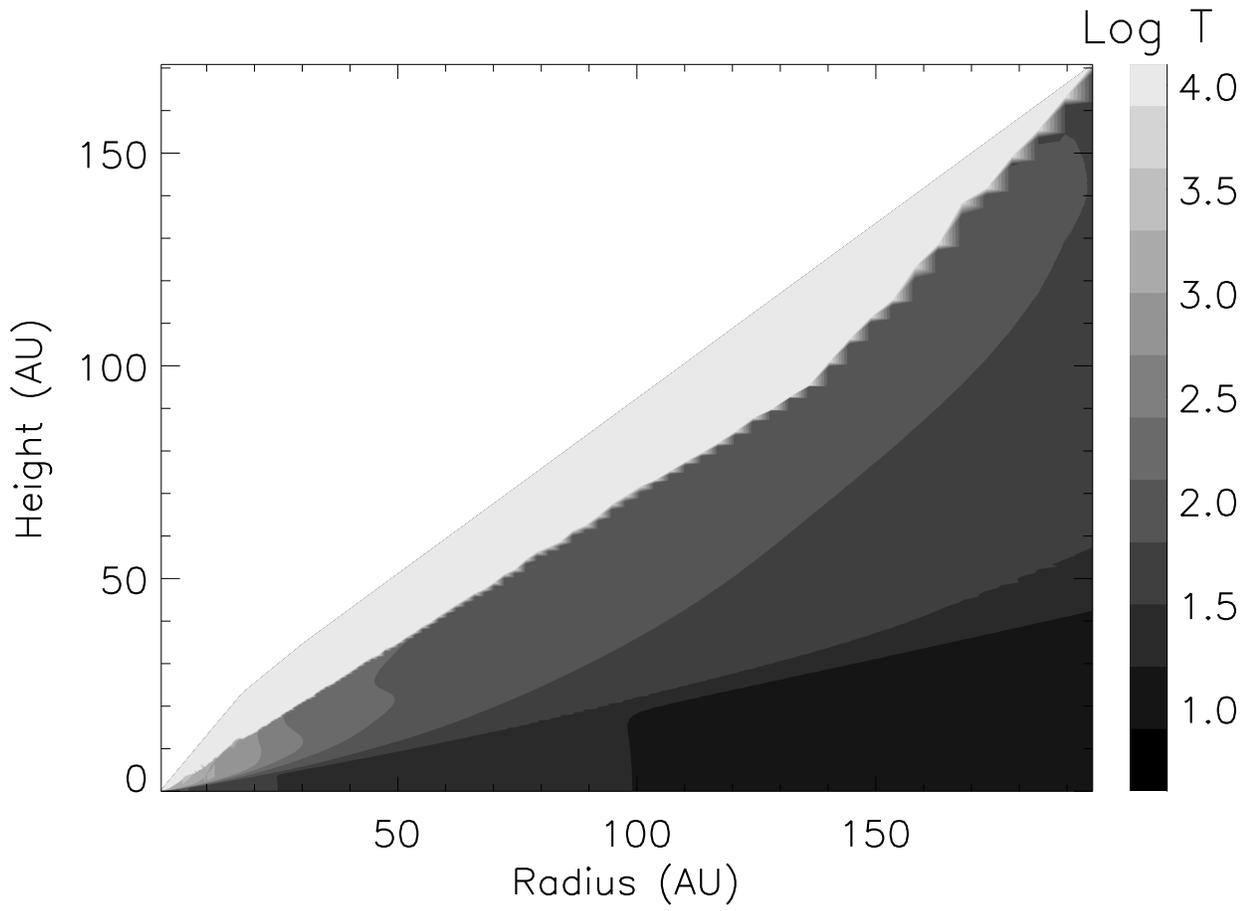}
\caption{Gas temperature contour plot of the fiducial model disk.}
\label{tempcon}
\end{figure}

\begin{figure}
\plottwo{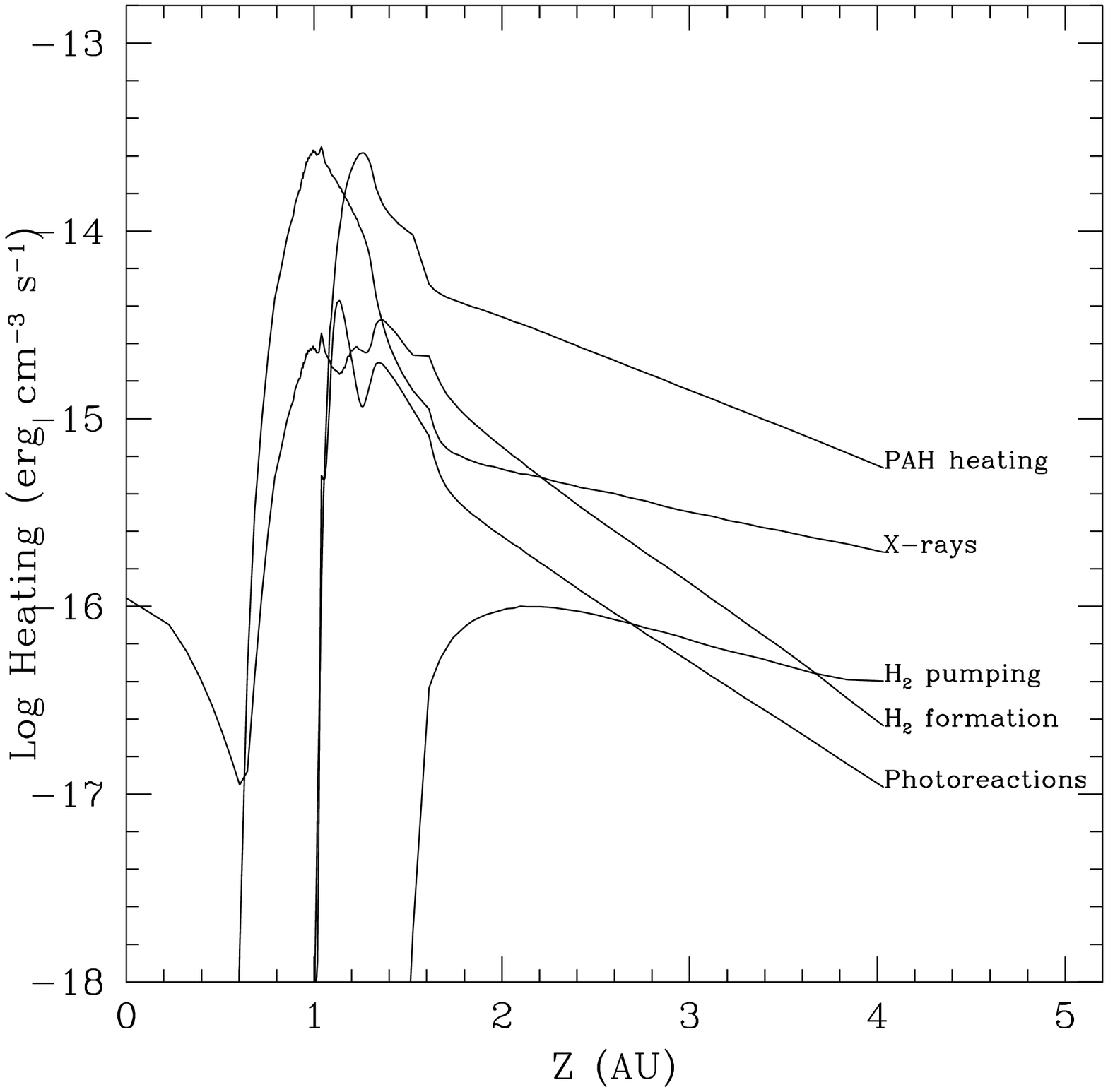}{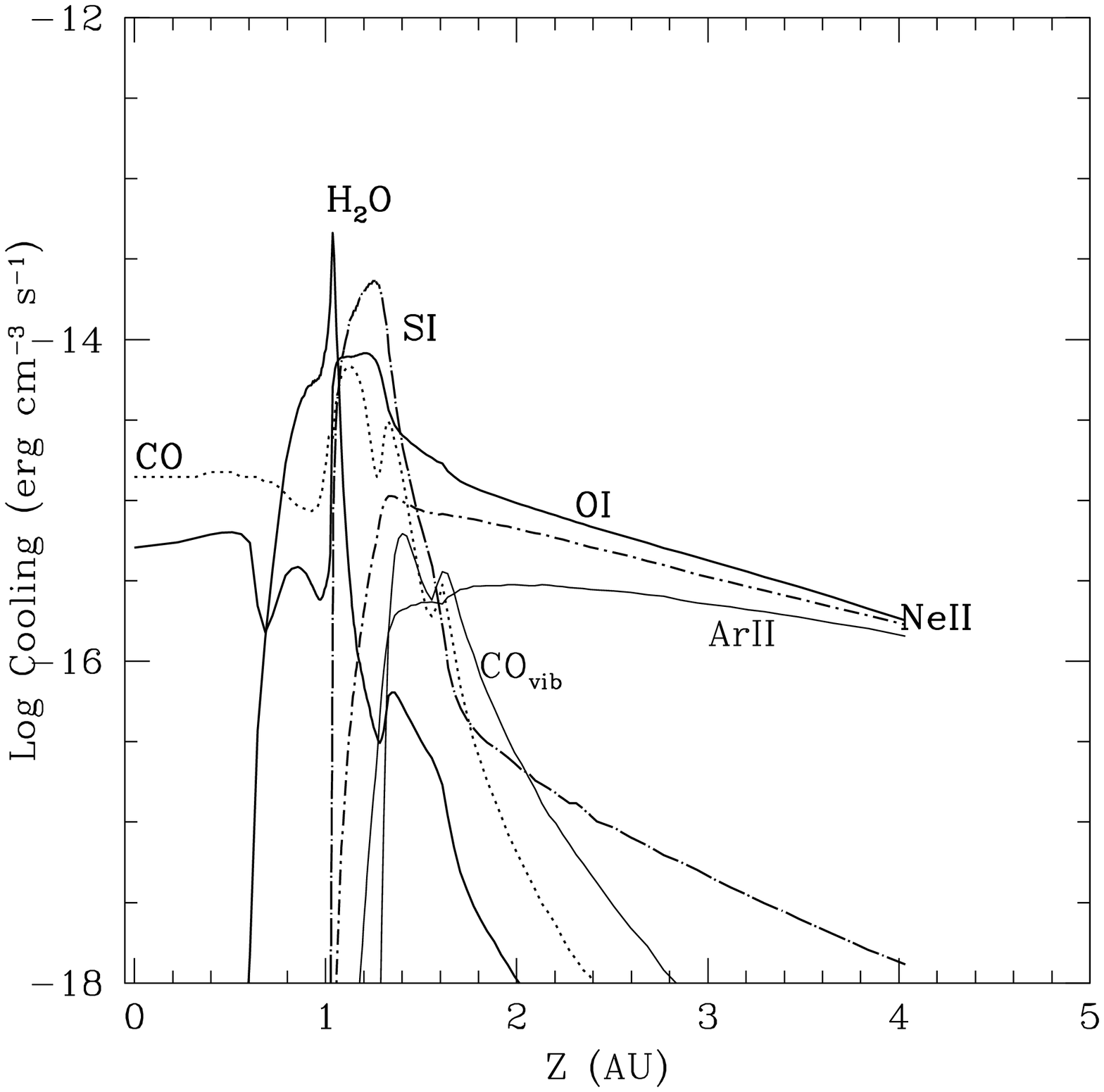}
\caption{The dominant heating and cooling agents in the disk at a 
radius of 8 AU, as a function of disk height. PAH heating here refers to
the sum of the heating by PAHs and by our grain size distribution,
although PAHs tend to dominate this sum. }
\label{thermal}
\end{figure}

\begin{figure}
\plotone{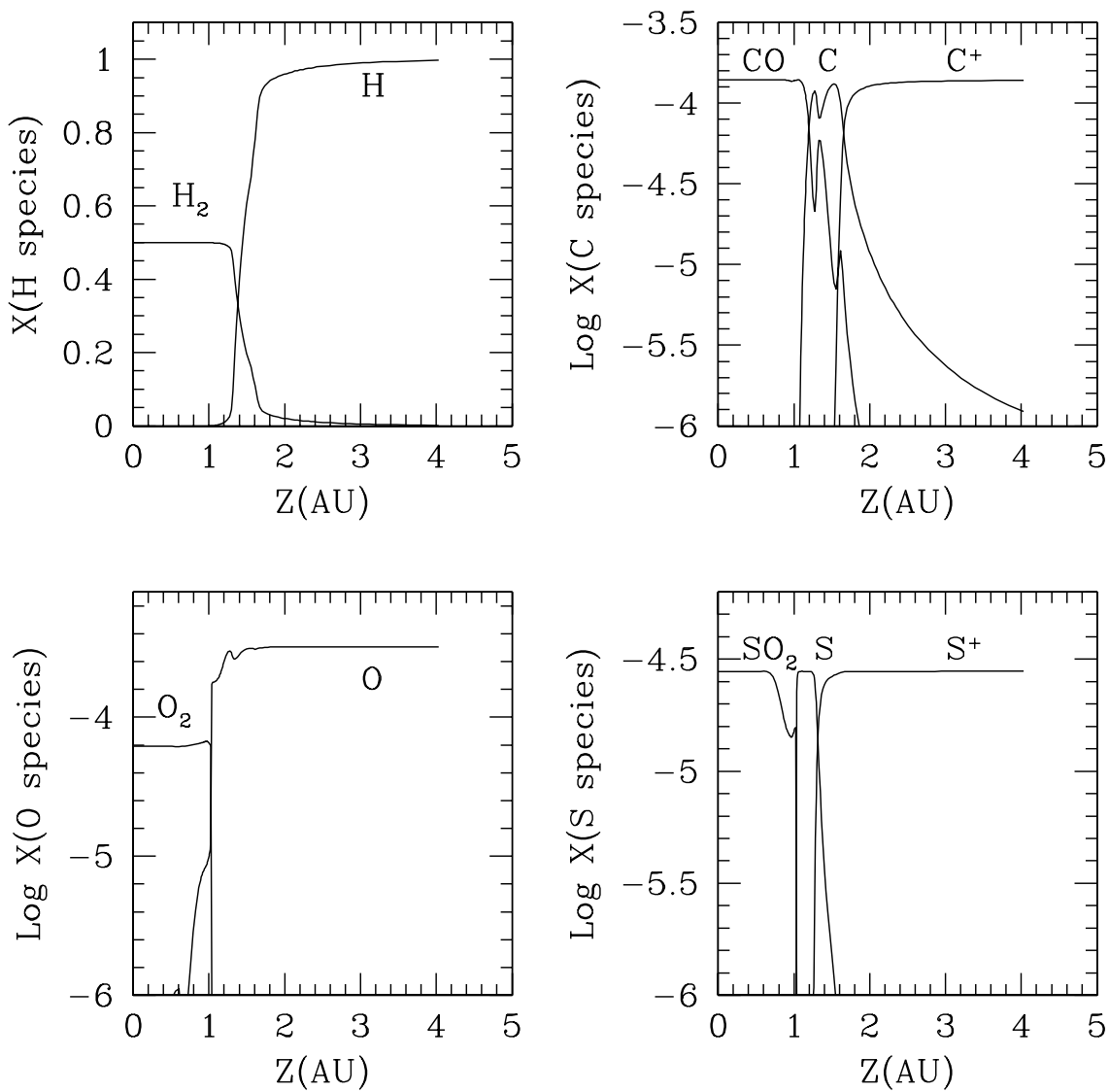}
\caption{The chemistry of hydrogen, carbon, oxygen and sulfur-bearing
species at a disk radius of 8 AU, with the abundances of dominant species
as a function of disk height. Note that midplane ($z\lesssim1$AU)
chemistry may be subject to formation of ices on grains (refer text). }
\label{fidchem}
\end{figure}

We describe the vertical structure of  a fiducial model of a  $0.03 {\rm M}_{\odot}$ disk
around a 1M$_{\odot}$ star.
The stellar and disk parameters are given in Tables~\ref{starpar}$-$\ref{fidpar}.  
The vertical density and temperature structure are shown as a function of
height at a distance of 8 AU from the central star in Figure~\ref{fidtemp}. 
Gas temperature contours  of the inner $r<10$ AU
region and the entire $200$ AU disk are shown in Figures~\ref{tempcon20} and \ref{tempcon}
respectively. Figures~\ref{thermal} and
\ref{fidchem}  depict the heating and
cooling and disk chemistry for the same vertical slice through the disk 
as Figure~\ref{fidtemp}. 

Figure~\ref{fidtemp} shows that the vertical gas and dust temperature structure
allows three distinct layers to be defined in the disk, (i) the midplane layer,
(ii) the intermediate layer and, (iii) the EUV-ionized surface layer. 
 In the midplane layer (where \av $\gtrsim$ 10 to the
central star) the gas and dust  temperatures are identical.  At \av $\sim 10-1$, gas and
dust temperatures deviate, to begin the intermediate layer.
Figure~\ref{fidtemp} also shows an ionized surface layer in the uppermost regions
of the disk at $r=8$ AU, where gas temperatures rise to $10^4$ K
at $z\sim4$ AU or \av \ $\sim 0.01$ from the neutral side of the IF. We will focus mainly
on the intermediate layers (between the midplane and the ionized layers of the disk) where
gas and dust temperatures are unequal,
because this is where infrared emission lines and FUV-generated
photoevaporative flows originate,
 generally at \av \ $\sim 10-0.01$ to the central star. 
 Figure~\ref{tempcon20}  shows the gas temperature
structure in the inner $r<10$AU where most of the mid-infrared line emission originates,
whereas Figure ~\ref{tempcon} shows the overall gas temperature structure for the entire 200 AU
disk. 

\paragraph{Midplane layer.}  In the densest midplane regions of the disk, densities and opacities are high at all radii  so that dust and gas are collisionally and radiatively coupled. 
 Here, heating and cooling of gas is dominated by dust collisions as the high densities allow efficient collisional transfer of thermal energy.  Dust and gas temperatures are almost equal and set by the nearly blackbody infrared radiation field.  
Other heating and cooling mechanisms are ineffective due to the large opacity (A$_{\rm V} \gg 1$) along the line of sight to the star and to the disk surface (Figs.~ \ref{fidtemp} and \ref{thermal}).   Due to the shielding of the disk interior,
steady-state, simplified chemistry results in high abundances of molecular species such as H$_2$, CO,
O$_2$ and SO$_2$ in the midplane (Figure~\ref{fidchem}).
However, note that the dust midplane temperature falls below the
freezing temperature of water ice for radii $\gtrsim $ few AU. Inclusion of  ice formation on
grains, presently ignored in these models,  may result in different midplane chemistry
and most of the elemental O may be incorporated in water ice, thereby depleting gas
phase CO, O$_2$ and SO$_2$ abundances, for example. Appendix B provides
a justification for our assumption of no freeze-out in the intermediate and EUV-ionized
surface layers and indicates conditions under
which freezing of species on cold dust grains may become important.
 Detailed chemical models of
disks which include ice formation have been developed by previous authors (e.g., 
Aikawa et al. 1997, Willacy et
al. 1998,  Aikawa \& Herbst 1999, Willacy \& Langer 2000, Markwick et al. 2002, Willacy 2007). 
However, as discussed earlier, we ignore ice formation as  the midplane regions are not important for either infrared line emission or for disk photoevaporation calculations, the two specific aims of our disk model calculations.

\paragraph{Intermediate layer.}
At intermediate layers, that begin where A$_{\rm V} \sim 10-1$ to the star,  gas and dust begin to thermally
decouple.  With increasing vertical
height and  decreasing opacity to the central star,   
dust grains are directly heated by stellar optical photons to higher 
temperatures (the ``superheated'' layer of the two-layer dust models). 
As \av\ to the star decreases, there is also increased
penetration of stellar high energy photons that heat the gas.   Grain photoelectric
heating (due to PAHs and small dust grains)
initiated by FUV photons, and  penetration of $\sim 1$ keV X-ray photons that directly heat the gas, result in higher gas temperatures. 
Physics and chemistry in the intermediate layer are dominated by stellar photons. 
Gas-dust collisions become less important at the lower densities in the intermediate
layer,  and cooling is mainly by gas line radiation  due to decreased opacity in the lines.
Therefore, it is here that the mid-infrared gas
 lines largely originate, and can be
observed in emission above the blackbody emission from the cooler midplane region. 
Figure~\ref{thermal} shows that at $r=8$ AU, much of the X-ray and FUV heating of the gas
is deposited at $z\sim1-2$AU, where \av $\sim10-1$. 
At this radius the [SI]$25\mu$m, [OI]$63\mu$m, 
[NeII]$12.8\mu$m and [ArII]$7\mu$m are strong emission lines. 
 The [OI]$63\mu$m line is one of
the main coolants over much of the disk  in the intermediate layer. 
 In the inner regions of the disk ($r\lesssim 20$ AU) gas temperatures due to X-ray heating are high and [NeII]$12.8\mu$m and [ArII]$7\mu$m transitions become important in the cooling of the
predominantly neutral gas ($T\gtrsim 1000$K) that lies just below the EUV-heated layer.  

Figure~\ref{fidchem} illustrates the chemistry of some of the important species at a disk
radius of 8 AU, and  shows that photodissociation of molecules and photoionization of atoms  
takes place in the intermediate layer. The H$_2$/H transition occurs where \av\ is $\sim 0.6$ to
the star, CO/C is at \av\  $\sim 1.6$, and C/C$^+$ is at  \av \  $\sim 0.4$. The density of gas at the H/H$_2$ transition is $\sim 10^7 {\rm cm}^{-3}$, with a locally incident stellar FUV field of
G$_0 \sim 10^5$, and FUV absorption is therefore dominated by gas opacity (e.g., Hollenbach \& Tielens 1999) mainly due to neutral carbon and H$_2$.  At the warm temperatures characteristic of this radius, H$_2$ is additionally destroyed by the chemical reaction H$_2 + $O leading to OH, which can be the dominant route at higher gas temperatures. The H/H$_2$ transition is therefore slightly deeper in the disk than the carbon ionization front. The conversion of C to CO takes place at higher \av\  (deeper) in the disk, as the location of this transition is determined by where  the dust opacity to FUV photons is $\sim 1$.  We also note that in disks with reduced dust (low values of $\sigma_{\rm H}$), lower formation rates of H$_2$ coupled with higher destruction rates due to the formation of OH for example, can lead to the formation of CO at higher $z$ and lower \av, even where the H$_2$ abundances are low. This result was also noted by Glassgold et al. (2004) who found in their X-ray heated gas disk models that the C/CO transition often was at higher $z$ compared to the H/H$_2$ transition. 
 
In the inner disk ($r\sim1$ AU) the high densities and strong FUV field make 
heating due to  H$_2$ formation and exothermic photoreactions  important in the \av $\sim3-1$ regions.
Grain photoelectric heating is also significant at these \av.
X-rays dominate the heating both at lower ($10 \lesssim {\rm A}_{\rm V} \lesssim 3$)
and at greater ($1 \lesssim {\rm A}_{\rm V} \lesssim 0.03$) heights.
[OI] is  the main coolant in the upper regions (${\rm A}_{\rm V} \lesssim 0.4$) of the intermediate layer.
  Deeper in the disk, where the densities are
high ($n\sim 10^9 {\rm cm}^{-3}$) and gas temperatures $\sim 1000$K, CO
vibrational cooling dominates ($ 1.6 \lesssim {\rm A}_{\rm V} \lesssim 0.4$). 
The CO/C and H$_2$/H transitions
are at \av \ $\sim 1.2$, with photodissociation by X-rays and FUV being important (e.g., Glassgold et al. 2004, Nomura \& Millar 2005).  At even higher densities and \av, dust collisions and
H$_2$O are the dominant coolants. 

In the outermost regions of the disk,  densities are low (since $\Sigma \propto r^{-1}$
and scaleheights are larger),
and a lack of collisional coupling can cause gas and dust temperature deviations to be significant. 
The effects of  gas heating by FUV and X-rays 
are less important far from the central source, and  gas does not get warmer than the dust until
higher values of $z/r$.  
Lower gas opacities result in efficient line cooling, gas temperatures
often drop below dust temperatures, and dust collisions can be a {\em heating} agent
in the intermediate layer.
 At $r\sim 100$ AU and at \av \ $\gtrsim 5$, gas and dust temperatures are equal. Above these
 heights, the gas temperature first drops below that of the dust.   Water cooling spikes at \av $\sim 4$, 
 but is generally unimportant. 
 In these intermediate layers,  X-rays and dust collisions
heat the gas up to heights where \av \ $\sim 0.4$, and CO rotational lines dominate gas
cooling in the regions $5 \lesssim {\rm A}_{\rm V} \lesssim 1$.  [OI]  takes over as
the main coolant at \av $\lesssim 1$.  Above \av \ $\sim 0.4$, grain photoelectric heating becomes significant,  and causes gas temperature to rise
above the dust temperature. The H$_2$/H transition and CO/C/C$^+$ transitions
occur high in the disk at \av $\sim 0.03$ and \av \ $\sim 0.5$ respectively.

\paragraph{The EUV layer.}
The uppermost layer in a disk is completely ionized by EUV photons from the central star. 
At $r=8$ AU, the EUV layer is situated at  a height $z\sim4.2$ AU, where the density
in the neutral gas just below the ionization front is $n \sim 2 \times 10^6 {\rm cm}^{-3}$
(Fig.~\ref{fidtemp}). 
The EUV layer at $r=1$ AU is at $z=0.25$ AU and $n \sim 3 \times 10^7 {\rm cm}^{-3}$,
while at $r=100$ AU, it is at $z\sim 70$ AU with the density $n \sim 4 \times10^4{\rm cm}^{-3}$
in the neutral disk just below.  
We numerically calculate the
ionization structure of the EUV layer  and predict line luminosities produced in this
layer in Hollenbach \& Gorti (2008).

\subsection{Line emission}
\begin{figure}[h]
\plotone{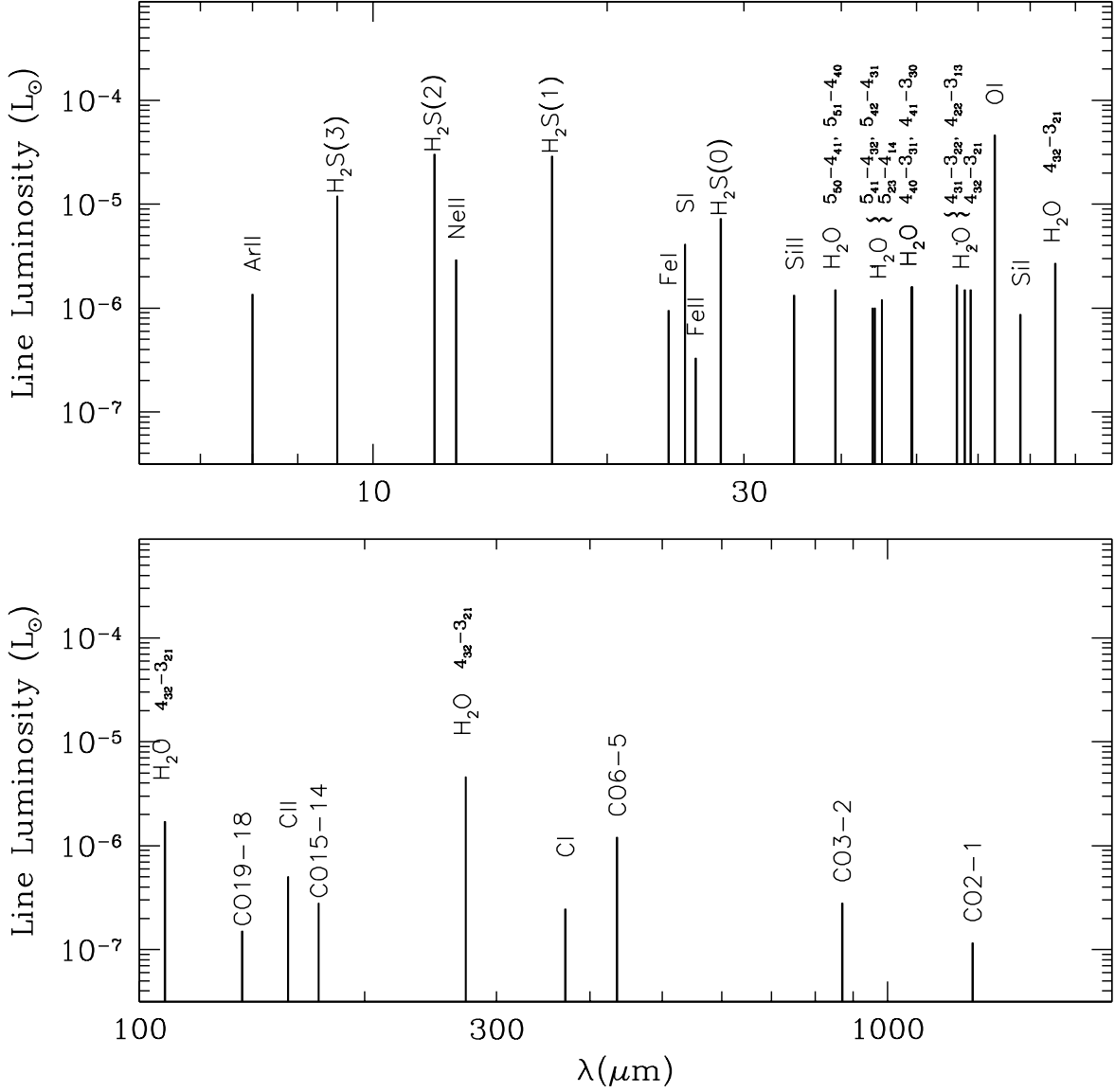}
\caption{Line luminosities as a function of wavelength for the fiducial disk model 
extended inward to 0.1 AU (i.e., disk is from 0.1-200 AU).
 Only a few strong water and CO  lines are shown in the figure.  
The  [OI]$63\mu$m line, S(1), S(2) and S(3) pure rotational lines of 
H$_2$,   and  certain H$_2$O  
lines  are some of the strongest emission lines from the disk.}
\label{spectrum}
\end{figure}

Gas  emission lines produced by dominant coolants can be important diagnostic probes
of the disk conditions. 
Strong atomic and molecular cooling in the intermediate layers of disks where gas tends to
be warmer than dust results in many mid-infrared emission lines originating principally from these layers. We discuss in this section the important transitions in the
mid-infrared, far infrared and sub-millimeter wavelength regions and
their detection with current and future observational facilities.  
These lines tend to be the most luminous lines arising from the $\sim 1-200$ AU disk, with
the exception of some optical, near-IR and mid-IR  lines produced in the very upper EUV-heated
layer (Hollenbach \& Gorti 2008) and some near-IR lines from thermally excited vibrational transitions of molecules in the inner ($r\lesssim 5 $ AU) disk.  Figure~\ref{spectrum}  shows 
the line spectrum from our fiducial disk of some strong emission lines in the infrared and
sub-millimeter wavelength regions. 
Our disk models are computationally very intensive because  a fine grid and many vertical
iterations are necessary
in order to resolve chemical transitions  and for convergence of the disk structure solution.
We therefore  restrict ourselves to a limited parameter survey of 
four different models of disks around a 1M$_{\odot}$ star to study the dependence of the
line strengths on  the stellar radiation field and on the dust/gas ratio. 
Table~\ref{fidline}
lists the predicted line luminosities from our models for some of the strongest
transitions. Model A
is our standard run (Table~\ref{fidpar}) which is also presented in Fig.~\ref{spectrum}.
 Model B has a dust opacity
($\sigma_{\rm H}$) 100 times lower
than Model A, and is considered to represent a more evolved disk that
has had substantial grain growth or settling. Model C is a case with no X-rays  and in Model D, the FUV
luminosity of the star has been increased from Model A by a factor of 10. 

In the following, we discuss each of the strong lines and compare our results with
model predictions by earlier authors when they exist. We also discuss our results
in the context of the ``Cores to Disks" (c2D) Spitzer Legacy Science Project data of gas line
emission from young, optically thick disks (Lahuis et al. 2007).  
 {\it Spitzer} IRS observations of disks by the c2D team were capable of detecting gas emission only from a few disks with very strong emission. 
The c2D survey of 76 circumstellar
disks detected [NeII]
emission from $\sim 20\%$ of sources, and [FeI]$24\mu$m emission from $\sim 9\%$. 
$8\%$ of the sources had H$_2$ 0-0 S(2) and/or S(3) line detections, and these
were anti-correlated with the [FeI]  detections. H$_2$ S(1) emission was seen in only
one source. Typical line luminosities range from $10^{-4}-10^{-6}$ L$_{\odot}$ for all the lines.
 We note that the many assumptions
inherent to the models regarding  stellar and disk properties 
make direct comparisons for each source possible only by individual detailed modeling. 
Here, in this first paper, we restrict ourselves to a more qualitative comparison of
results with data, and model only one source, TW Hya, in detail.

\subsubsection{H$_2$ rotational lines} 
The H$_2$ rotational lines arise from the inner disk, at $r\lesssim 20$ AU,
 and at heights where \av \ to the star ranges from $\sim 1-3$ at the inner edge to
$\sim 0.3$ at larger radii.  
The dominant H$_2$ line in a region 
shifts as the temperature in the \av $\sim 1$ region drops with radius, 
 from the 9$\mu$m S(3) and 12$\mu$m S(2)  lines (at T$\sim 500-1000$K) in the inner
disk to the 17$\mu$m S(1) (at T $\sim 100-900$K) and the 28$\mu$m  S(0) lines
(at T$\sim 60-200$K) from the $5-20$ AU region. 
The S(3) and S(2) line emission peaks mainly
at the inner edge (or rim, $r \lesssim 1$ AU)) where the gas temperatures are  $\sim 1000$ K
at these heights.  S(1) emission  is also significant here ($30\%$
of total luminosity),  but most of the  S(1) and all of the S(0) line
emission is from  $r\sim5-20$ AU.  
The S(1) line at $17\mu$m is the strongest line
for all our disk models,  where we only report line emission from beyond 0.7 AU. 
Since most of the S(2) and S(3) emission arises from the inner edge of the disk in 
Model A, we calculated an additional case with a smaller inner disk radius. 
The S(2) and S(3) line strengths are nearly doubled when the gas disk is extended 
closer to the star (to 0.1AU),
or if we consider direct illumination of the rim of  a disk with
an inner hole (i.e., we include the $0.5-0.7$ AU region in our line
emission).  In these cases, the S(2) line is comparable in strength to the S(1) line.

 Gas heating in the H$_2$ emitting region is mainly
due to FUV-initiated grain photoelectric heating by PAHs,  X-rays and
FUV  pumping of H$_2$. 
In Model B,  where the dust and PAH abundances are  lowered by a factor
of 100 from the standard Model A,  gas heating due to the photoelectric
effect is reduced and the  decreased heating leads to  lower gas
temperatures. This is especially true in the $r \lesssim 20$ AU region.  Here
PAHs dominate the heating in the \av $\sim 0.1-1 $ region and
X-rays  dominate at slightly higher \av  \ (Fig.~\ref{thermal}).  Gas-dust
collisional cooling is important in the \av $\sim 0.1-1$ region, and is
lower relative to other coolants for Model B due to the lower $\sigma_{\rm H}$. 
The S(2) line arises from higher $z$  (and lower \av) where PAHs heat, and 
 its strength decreases (at $r<5$ AU)
 due to the lower gas temperature.   On the other hand, S(1) emission increases here,
which compensates for its decrease at greater radii ($5-20$AU),
and the S(1) line strength remains relatively unaltered. 
The S(0) line is  unchanged as it  arises at greater \av \ in the
$5-20$ AU disk where X-rays dominate heating and 
where the gas temperature is marginally affected by the reduced PAH abundance.
Because of its lower excitation energy, the S(0) line is also less
temperature sensitive than the S(1), S(2), and S(3) lines. 
The effect of neglecting X-rays (Model C) removes this heating agent, and
the resulting lower  temperatures  decrease all H$_2$ line strengths
in the disk.  An increase in the FUV luminosity (Model D) by a factor 
of 10 from the standard model increases the strength of the lines as 
is to be expected. FUV heating in this case is important even at the higher
\av \  layers of the $r<20$ AU disk  region where the S(0) line originates.  

The rotational line luminosities of H$_2$ are  $\sim 10^{-5}$ L$_{\odot}$, 
well above the sensitivities of   {\it Spitzer} and in favourable
cases (luminous UV and X-ray stars that are nearby), 
ground-based facilities such as the TEXES spectrograph (Lacy et al. 2002)
or MICHELLE (Glasse et al. 1997) on large telescopes.  However, the high dust continuum
at these wavelengths ($\sim 10^{-3}$ L$_{\odot}$ for a resolving power
of 700 provided by the IRS on  {\it Spitzer}) indicates a very poor
line-to-continuum ratio for  {\it Spitzer}, and instruments with both high
resolving power and high sensitivity are needed to see the lines. 
  For very young disks around cTTs, with strong FUV heating due
to perhaps a higher accretion rate (e.g., Gulbring et al. 1998), the
H$_2$ lines are strong, and the S(2) line has been detected  from several sources
 by the 
c2D team with the Spitzer Space Telescope.  
Our predicted H$_2$ S(2) line strength for a  cTTs disk
compares reasonably well with the typical observed value of $10^{-5}$ L$_{\odot}$
(Lahuis et al. 2007).
For computational tractability,
we have restricted our disk models to radii larger than $\sim 0.7$ AU. Extending
our models to smaller inner disk radii results in thermal emission
lines at higher excitation energies, such as the S(2) and S(3), being stronger. 
The strength of the S(0) and S(1) lines are comparable to the S(2) and S(3) 
lines in our models. However, S(0) and S(1) are
 undetected by the IRS, and we suggest that the decreasing
line-to-continuum ratio at longer wavelengths is perhaps the explanation
for the non-detection.  The c2D survey upper limits for the S(0) and S(1) lines 
are $\sim 10^{-5} {\rm L}_{\odot}$ consistent with our model predictions. 
The EXES instrument on SOFIA has a sensitivity
similar to that of the IRS, but a higher resolving power  (R $\sim 10^4$)
and thus is better suited to detecting lines in the presence of a strong continuum.  It is
capable of detecting lines with a luminosity of $\sim 3 \times 10^{-6}$ L$_{\odot}$ for a source at the distance of Taurus (150 pc)  and will probe gas in the $1-20$ AU regions of nearby disks.  However, 
limited sensitivity  will make the detection of gas emission lines  from distant disks challenging. 
 
Our H$_2$ S(1) and S(2) emission line strengths are 
in agreement
with the results of Nomura \& Millar (2005), who obtain line luminosities
of $\sim 10^{-5}-10^{-6}$ L$_{\odot}$ for the H$_2$ S(1) and S(2) lines in their
disk model with UV heating.  Our results indicate that the FUV and X-ray luminosities of
the central star are important  in determining disk H$_2$  line emission,
and that the dust properties (settling, coagulation) have a smaller influence. 
 
\subsubsection{[NeII], [NeIII] and [ArII]}
The fine structure lines of [NeII], [NeIII], and [ArII] originate from high-temperature
gas (T $\sim 10^{3}-10^4$ K) heated and ionized by X-rays and  EUV photons from the central star. 
 We discuss
the emission from the EUV region in a separate paper (Hollenbach \& Gorti 2008)
and here present only the calculations of [NeII] and [ArII]
line emission from the X-ray heated neutral
gas just below the completely ionized EUV region (\av $\sim 0.02-2$).
X-rays ionize Ne and Ar in the neutral disk and the ionization fraction
is  $\sim 10-20$\%  in the emission regions (see also Glassgold et al. 2007).
  Radially, [NeII] and [ArII] emission from the X-ray  heated gas 
  is limited to $\lesssim 20$ AU,
as first found by Glassgold et al. (2007). Beyond  20AU the gas temperature
falls significantly below the excitation temperature,  $\Delta E/k \sim 1100$K, for
[NeII]$12.8\mu$m. 
   At the very surface of the neutral region, close to the
EUV-heated region where the gas is fully ionized, and at the inner edge of the
disk, [NeIII] is abundant. 
One important effect, noted by Glassgold et al. 2007, is the charge exchange reaction
of Ne$^{++}$ with H, which lowers the Ne$^{++}$ abundance and thus the
strength of [NeIII].
 [NeIII] luminosities from the X-ray heated 
gas are therefore low and  typical [NeIII]/[NeII] line ratios are $\lesssim 0.1$.
[ArII] arises from higher temperatures (heights) than the [NeII] line.
We  find that [NeII]  can contribute significantly to the total
cooling of gas in these regions, at \av $\lesssim 0.1$ within $\sim 20$ AU. 

Predicted [NeII] and [ArII]  line luminosities are moderately high $\sim$ few $10^{-6}$
L$_{\odot}$ for all our disk models, and [NeII] has been observed
around many young disks to date (Uchida et al. 2004, Geers et al. 2006,
Lahuis et al. 2007, Pascucci et al. 2007, Espaillat et al. 2007).  Calculated and
observed [NeIII] luminosities are lower than the [NeII] luminosity
by a factor of $\sim$10.  However, X-ray produced [NeII] fluxes
often fall short of the observed line fluxes, suggesting a contribution  from the
ionized EUV-heated layer as well. Since we ignore the EUV-heated layer, 
Model C with no X-rays
shows no emission from these lines.  However, dust properties and 
FUV luminosity in the presence of X-rays do affect the line strengths indirectly via their effect
on the overall disk structure.  Decreased dust opacity (or grain growth) 
results in a more ``settled'' disk with a smaller flaring angle,
reducing the intercepted X-ray flux at the surface and thereby the
[NeII] and [ArII] line emission. On the other hand, a higher FUV field (Model D)
produces higher gas temperatures at the surface and greater disk flaring.
The X-ray heated surface layer in this case subtends a larger solid angle. 
The increase in X-ray photon penetration in combination with the extra
FUV heating of the gas results in  higher line luminosities.

Our Model A run is similar to the standard model of Meijerink et al. (2007),
and the [NeII] luminosities are in good agreement. Meijerink et al (2007) do
not include cooling by [NeII] which we find is important for the disk
structure. Consequently, their gas temperatures in the surface [NeII] emitting layer
are in general higher, which should produce higher [NeII] luminosities.
However, they use a dust-determined density distribution that produces a surface
subtending a smaller solid angle, and this decreases the
amount of X-ray flux absorbed by the disk and lowers the [NeII] luminosity. 
Thus, the two effects partially cancel and their results are similar to
our Model A. 

To further study the differences between our model and that of Meijerink
et al. (2007),  we compared our model
results with their results for a disk with no accretion heating, using an
almost identical model disk. In this  run, we let the density distribution
be determined by the dust temperature structure (as in their model) and used their assumed
dust properties. We find that our [NeII] luminosities are, in fact, a factor 
of $\sim 4$ lower.  This difference shows the importance of [NeII]
and [ArII] cooling on the gas temperature.

We note that the inclusion of  [NeII] emission from the EUV-heated region would enhance the predicted total [NeII] line luminosities of all the models. For a typical EUV  luminosity from a solar-type star of $\phi_i = 4 \times 10^{41} {\rm s}^{-1}$, the calculated [NeII] line luminosity from the ionized HII-region above the disk is $\sim 10^{-6} {\rm L}_{\odot}$. The [NeIII]/[NeII] line ratio depends on the EUV spectrum  and is higher for stars with a hard EUV spectrum (Hollenbach \& Gorti 2008).

\subsubsection{[FeI] and [FeII]}
[FeI]$24\mu$m  and [FeII]$26\mu$m  line emission can be significant
and detectable,  even though they are  
seldom  important disk coolants.  [FeI] emission originates relatively deep at
 a vertical height given by \av$\sim3-10$ to the star, and
 at $ r \sim 4-30$ AU.  At these vertical heights the dust and gas temperatures
begin to decouple and the gas  ($T_{gas} \sim
100-200$K) gets warmer than the dust.   As the \av \
to the star decreases with height, Fe is rapidly ionized to form Fe$^+$. 
[FeI] line luminosities are $\sim 10^{-6}-2\times10^{-5}$ L$_{\odot}$ in all our disk models.
[FeI] arises from deeper layers and is strongly affected by the disk
opacity (Model B). Decreased dust opacity in the disk increases the 
density at the \av $\sim 5$  layer whereas the gas, heated mainly by X-rays,
is at the same temperature as in Model A, and the line strength increases.
The [FeI] luminosity increases
by a factor of 20, to $2 \times 10^{-5}$ L$_{\odot}$, with a reduction in dust
opacity by 100.   With no X-rays incident on the disk (Model C), the [FeI]
luminosity decreases due to the reduced heating. However, closer
to the surface and the Fe/Fe$^+$ ionization front, there is 
some heating due to PAHs in this model which raises the gas temperature
slightly above that of the dust, to produce [FeI] emission. 
An increase
in the FUV field (Model D) results in higher gas temperatures, but on the
other hand, also causes photoionization to Fe$^+$.  [FeII]  increases with the
FUV luminosity of the star and can be stronger than the [FeI] line (Model D).

Our predicted [FeI] luminosities for all the cases considered
here are at least a factor of 5 lower than the most [FeI] luminous {\it Spitzer}
 IRS detections of the c2D team (Lahuis
et al. 2007), where the  [FeI] luminosity when detected is $\sim 10^{-6}
-10^{-4}$ L$_{\odot}$.
From our analysis, we find that [FeI] emission is strongest when
the disk dust opacity is very low, as in Model B.   Further drastic reductions
in dust opacity make the disk only marginally optically thick and the two-layer
dust radiative transfer is no longer applicable. We note that
our earlier models of optically thin disks predicted enhanced
[FeI] emission from massive disks with low values of $\sigma_{\rm H}$ 
(Gorti \& Hollenbach 2004). 
These results are all suggestive of increased dust settling or grain growth
enhancing the gas/dust ratio and decreasing $\sigma_{\rm H}$
in the $r\lesssim40$ AU regions of disks where [FeI] is observed.  
We  also note that our assumed gas phase abundance of Fe ($10^{-7}$) may
be an underestimate in disks. We will address these  issues in  future work on the
modeling of the observed  [FeI] emission in disks. 

\subsubsection{[SI]} 
The [SI]$25.2\mu$m line  arises from the $\sim 5-20$  AU region in disks,
at \av $\sim 0.5-3$. This line is strong  in our model disk,
although it is often not the dominant coolant in the region where it arises.  In the
sulfur emitting radial regions of the disk,  heating is mainly due to X-rays and
FUV photons, and the main coolants are CO, [OI], [SI] and dust collisions. 
We include detailed sulfur chemistry in our models, and
assume a gas phase abundance typical of the diffuse interstellar medium,
$2.8 \times 10^{-5}$ (Savage \& Sembach 1996).
  Sulfur is molecular (SO$_2$) in the interior of the
disk,  is photodissociated to form S in the \av $\sim 1-3$ region 
(column density $\sim 10^{22}$ cm$^{-2}$ for our fiducial case), and is photoionized to form
S$^+$ in the upper layers of the disk (\av $\lesssim 0.3$).  [SI] emission is reduced slightly for the reduced
dust opacity disk (Model B) due to decreased heating by PAHs.
As X-rays are an important heating mechanism in these regions,
the no X-ray model (C) shows a reduction in the [SI] line luminosity as is
to be expected.   Increasing
the FUV field of the central star increases the heating and raises the gas
temperature to increase [SI] emission in Model D. 

Lahuis et al.(2007) do not detect [SI] emission from their disks, though 
they do detect the H$_2$ S(2) line, which we find to be comparable in strength.
This could be because of the rising continuum from dust at the longer
wavelength of the [SI] line, making line detection difficult.  Sulfur may also deplete
on grains in the emitting region, decreasing line emission. 
Watson et al. (2007)  report the presence of strong [SI]$25\mu$m emission from two
class I protostellar disks, but this emission may arise from shocks
in the outflows around the protostars. We also note that
our [SI] line luminosities are higher than predicted by Meijerink et al. (2007).
They assume a much lower gas-phase sulfur abundance,  and the trace amounts
of carbon not locked in CO allows the formation of CS  
in their models in the regions where [SI] arises,
thereby reducing the atomic S abundance. In our models the elemental S abundance
far exceeds the {\em available }carbon abundance so that the formation of CS does
not appreciably deplete atomic sulfur.  Further,
they do not include the effects of FUV radiation, which we find very important
for determining the chemistry (ionization potential of S is 10.36 eV), and
also for gas heating. Our gas temperature in this intermediate layer is 
typically higher than their models due to the inclusion of FUV.

\subsubsection{[SiII]}
[SiII]$34.8\mu$m emission peaks slightly further out in the disk than many other mid-infrared
emission lines which tend to arise mostly from the region inside of $25$ AU.
[SiII] arises from regions between 
5 and 40 AU. It is relatively unaffected by changes in dust/gas ratio and
X-rays, but increases significantly with FUV luminosity due to the combined
effects of higher gas temperatures and a greater abundance of ionized silicon.
This line falls in a relatively insensitive wavelength region of the IRS spectrometer
on {\it Spitzer}, and also at a wavelength with high dust continuum emission, and is therefore
difficult for {\it Spitzer} to detect. 

\subsubsection{[OI]}
The [OI]63$\mu$m line is often the strongest cooling transition in disks, and because of its low excitation temperature ($232$ K) can arise from the surface regions of almost the entire disk. 
 [OI] emission originates in the atomic
layer at \av $\sim 0.1-3.0$ at all radii. 
The luminosity per logarithmic interval ($d{\rm L}/d \ln r$) rises nearly linearly out to $\sim 20 $ AU, 
flattens beyond $20$ AU and then declines very slowly from
$\sim 120 $ AU. Because the line tends to be optically thick,
the line luminosity is effectively determined by the gas temperature at
the photosphere of the line, which typically occurs
at \av $\sim 1$. 
In the regions where [OI] originates heating is due to both FUV photons
 and X-rays, and lowering the dust opacity (Model B) or the X-rays (Model C)
only marginally decreases the line luminosity because the other mechanisms
take over the heating. 
 In Model D with higher FUV,
there is more heating, the gas temperature is higher and [OI] emission
is higher. Our calculated
[OI] luminosities are high, $\sim 10^{-4}$ L$_{\odot}$ 
and this line  is therefore a promising probe of the presence of atomic 
 gas in the outer regions of disks.
Our model results are also consistent with the results of Meijerink et al. (2007) who obtain an
[OI]$63\mu$m  line luminosity of $\sim 5 \times 10^{-5}$ for their standard model. 
 The PACS spectrometer on  {\it Herschel} is sensitive enough to detect OI emission from disks at 150 pc (e.g, Taurus, Chameleon and $\rho$ Oph star-forming regions). PACS has relatively modest resolution and will not be able to resolve the line. 
The [OI] $145\mu$m line is  10-100 times weaker than the $63\mu$m line, but may still be detected by PACS from disks around FUV-luminous sources.   ISO detections of of the [OI] lines have been reported by Creech-Eakman et al. (2002)
from the disks around AB Aur and other sources, 
and the observed luminosities $\sim 10^{-4}$ L$_{\odot}$ agree well with
our model results. 

\subsubsection{[CI] and [CII]}
[CI] and [CII] are marginal coolants in the disk and increase in luminosity
as the disk radius (and hence emitting area)  increases.  
They originate where CO dissociates,
and in general the C column density in the outer disk is low($\sim 2\times 10^{16}
{\rm cm}^{-2}$ at  $r\sim 100$ AU), because the C is readily
photoionized to C$^+$. [CII] emission is stronger than [CI], and is
high when the FUV luminosity is high (Model D) due to the higher
gas temperatures and C$^+$ column densities in the outer disk.  
Our model results agree with the
predictions of Kamp et al. (2006) and Meijerink et al. (2007) in their
line luminosities, when we truncate the disks to match their assumed
outer radii.  The [CII]$158\mu$m transition, often a strong emission line along
with  the [OI]$63\mu$m line  in photodissociation regions (e.g., Kaufman et al. 1999), 
is much weaker than [OI] in disks because of their  higher densities and temperatures. The [CII] line
may  be 
too weak for detection by HIFI on  {\it Herschel}. The sensitivity of the HIFI instrument is about 10 times lower than that of PACS, and with the CII line luminosities  calculated as being typically   $\sim 10^{-6}-10^{-7}$ L$_{\odot}$, detecting [CII] emission from disks will be difficult, except for very nearby sources.

\subsubsection{CO rotational lines}
In the outer disk, CO is the main coolant from \av \ $\sim 5$ where the gas temperature
begins to deviate from the dust temperature, up to \av \ $\sim 0.5$ where the CO is
photodissociated. Often in this region, the gas temperature is less than the dust
temperature, and thermal balance is achieved when the heating of gas by 
collisions with warmer dust  equals the CO rotational line cooling. 
We also find that  CO emission originates from heights where it is readily
desorbed as ice from grains and is not significantly affected  by freeze-out (Appendix B). 
In these intermediate layers, a reduction in dust opacity (Model B) 
marginally decreases
the low J rotational line emission due to the lower gas temperatures. 
Low J rotational lines are
relatively unaffected by X-ray heating, as X-rays are 
only important at depths greater than 
those characteristic of the surface layers producing the CO lines.
 An increase in the stellar FUV flux (Model D) can 
heat the gas in the outer disk (increasing line luminosities) but also 
increases the photodissociation rate of CO molecules, moving the CO
surface to lower heights where gas temperatures are lower.  We find that
increasing the FUV field by a factor of 10 does not appreciably change the 
line luminosities. 

Ground-based CO observations with the JCMT, SMA and IRAM receivers
are capable of detecting gas  line luminosities of $\sim 10^{-8}{\rm \ to 
\ }10^{ -9}$ L$_{\odot}$ from sources at $\sim 100$ pc.
Our calculated low J CO line luminosities are much higher than these
limits, and for nearby disks
that are not significantly beam diluted,  CO is a very good indicator of the presence
of molecular gas in the outer ($\gtrsim 20$ AU) regions. 

\subsubsection{H$_2$O lines}
We estimate the strengths of  H$_2$O  lines using the derived density and temperature structure and solve for the level populations (lower 167 levels) after a disk solution for the temperature, density, and chemical abundance structure has been obtained using analytical approximations for the total cooling by H$_2$O molecules (GH04).  We note that cooling transitions are often optically thick, and we use the escape probability formalism in calculating the line strengths as described in Hollenbach \& McKee (1979).

Water is a very important coolant in the disk in the lower parts of the
intermediate layers where the dust
and gas temperatures begin to deviate (i.e., vertical heights where \av $\sim 10-1$) and where the
gas temperatures are $\sim 100-1000$ K.  Gas phase water abundances in the shielded
interior of the disk are low, $\sim 10^{-8}$ -$ 10^{-6}$, where H$_2$O formation is
initiated by the weak ionization of H$_2$ to form H$_2^+$ which then reacts with H$_2$ to form H$_3^+$. The H$_3^+$ reacts with O to form OH$^+$ and subsequent reactions with H$_2$ ultimately leads to H$_3$O$^+$, which recombines with electrons to form H$_2$O.  At higher
temperatures, typically $\gtrsim 300$K, the main formation route
is by neutral-neutral reactions of O with H$_2$ and OH with H$_2$. The high
temperatures are needed to overcome the large activation barriers 
of O and OH leading to H$_2$O, and attained typically close to the \av =1
 layer of the inner ($\lesssim 40$ AU) disk, where the hydrogen is still
molecular. Water emission is strongest near the H$_2$ dissociation front,
where gas temperatures and H$_2$ abundances are both high, and water
is a strong coolant here (Fig.~\ref{thermal}). Line emission in the
dense ($n \sim 10^{8} - 10^{11}$ cm$^{-3}$) gas peaks as a function of $z$
 where H$_2$ and H$_2$O 
begin to photodissociate and gas temperature is rising.  
Although the
total cooling is high ($\sim 3 \times 10^{-5}$ L$_{\odot}$), the emission is distributed
over many different transitions at various wavelengths so that individual strong lines
tend to be typically $\sim 10^{-6}$ L$_{\odot}$. 
We note that in dense
regions of disks, where the dust grain temperatures may fall below $\sim 85$K,
water may freeze out on dust grains and may be depleted from the gas phase 
(see Appendix B). As we ignore freezing in our models, our predicted line
luminosities may be upper limits for some of the longer wavelength and low-lying
transitions of water. As shown in Appendix B, freeze-out of water may become
important at large disk radii and at deeper \av \ (where gas temperatures and photodesorption rates decline)
where these longer wavelength emission lines originate.
 Table~5
lists some of the strongest infrared emission lines, many of which are detectable
by HIFI and PACS on  {\it Herschel}, by SOFIA (if atmospheric transmission permits),
 and a few mid-IR lines may be strong
enough to be detected by   {\it Spitzer} IRS.  Line luminosities of transitions that may be affected by
freeze-out are indicated as upper limits. 
 The ortho-H$_2$O $3_{12}-3_{03}$
line at $274\mu$m, accessible by HIFI,  is the strongest emission line and arises from the
 hot, dense gas at the inner edge of the disk ($r\lesssim 1$AU). 
  H$_2$O line emission increases 
for lower dust opacities, because of higher gas densities at similar
\av, but only by a factor of $\sim 3$ for a reduction in $\sigma_{\rm H}$ by 100 
(see Table~\ref{fidline}). In the disk with no X-rays, gas heating and destruction of H$_2$O are both
lower, and the emission region shifts to higher $z$ and slightly lower \av\  where FUV  heating
is strong (see Fig.~\ref{thermal} for the Model A disk).  The warmer gas at
lower \av\ but lower H$_2$O abundances act in opposition to keep the total
water cooling nearly the same as the Model A disk.    Increasing the FUV field (Model D)
has similar effects of raising the gas temperature but simultaneously increasing
H$_2$O photodissociation rates, and the total water cooling is not much affected.

\section{Modeling the disk around TW Hya}
We apply our disk models to explain observed gas emission from the disk around the nearby, well-studied star, TW Hya.  The properties of the central star and those of its dust disk are well determined,
and therefore  the number of free input parameters to our model are greatly reduced.  Rotational CO emission line data from three different transitions have recently been obtained using the SMA
by Qi et al. (2004, 2006) and we compare our model results with this uniform dataset which probes gas at different temperatures (and locations) in the disk. 
We also compare our results with the observed [NeII] emission by the  {\it Spitzer} IRS
(Uchida et al. 2004). This source is therefore well-suited for demonstrating the applicability of our models to determine the physical conditions and constrain the spatial location of the gas. 

 TW Hya is the closest (51 pc, Mamajek 2005) known young T Tauri star (age
 $\sim 4-10$ Myrs, Webb et al. 1999),
  and has a massive, face-on optically thick disk.
  TW Hya is known to be a strong X-ray source (Kastner et al. 1997). 
   Muzerolle et al. (2000)
 estimate an ongoing accretion rate  of $4 \times 10^{-10}
 {\rm M}_{\odot}/$yr  and the star has an observed FUV spectrum (IUE, Valenti et al. 2000). 
TW Hya is also known to have a strong Lyman $\alpha$ emission  component in its FUV flux (Herczeg et al. 2004). Ly$\alpha$ excites the higher ro-vibrational levels of the ground electronic state of H$_2$  and populates higher electronic states. This is an important effect in determing the excited electronic and ro-vibrational emission of H$_2$ from the gas in the disk (Nomura \& Millar 2005). For the largely thermally excited pure rotational emission discussed in this paper, the stellar Ly$\alpha$ component is expected to be marginally important. For simplicity, we ignore the Ly$\alpha$ emission from the star.  
The dust disk has been extensively modeled by Calvet et al. (2002) and  extends from $\sim 4-200$ AU in size (Hughes et al. 2007). 
Assuming a gas/dust mass ratio of 100, Calvet et al. determine the total disk mass (gas+dust) to be $\sim 0.03$M$_{\odot}$, if they 
take a  maximum dust grain size of 1 cm.
  Table~\ref{twhyapar} lists the observational data and other input parameters used to model the gas emission.    
  Our model disk is constructed by using the dust disk model of Calvet et al. (2002) and by adding a gas component with an interstellar gas to dust mass ratio ($\sim 100$). We model the gas (and dust) components and calculate the density, temperature and chemistry as a function of spatial location in the disk. 
We compute the line luminosities of the CO and [NeII] lines and compare them with observations in Table~\ref{twhyalin} for two different disk radii. Model I has a radial extent of 174 AU, similar to that of the dust disk (e.g., Roberge et al. 2005) and Model II is a truncated gas disk with a radial extent of 120 AU.  

The [NeII] luminosity in the two models agrees extremely well with the observational
[NeII] luminosity.  In both models, the [NeII] line originates from the inner rim at 4 AU from X-ray heated gas exposed directly to the central star and
so there is little dependence on the outer disk radius. Though there appears to be good agreement between theory and observations, this result should be taken with caution.
 The inner rim is a region subject to flows, viscous accretion and photoevaporation (e.g., Chiang \& Murray-Clay 2007, Alexander et al. 2006) and is not treated very accurately in our steady-state models. 
 
 The CO lines  in Model I are stronger than observed, a result also obtained by Qi et al. (2006) in their disk modeling.  In our Model I, we overestimate the CO lines by a factor of $\sim$ 2. 
 The CO 2-1 and 3-2 lines originate from the entire (174 AU) disk, whereas the CO 6-5 line is emitted by slightly warmer gas confined to within $\sim 100$ AU from the star.  The truncated disk Model II appears to better match observations, although we still overpredict the CO 6-5 line by a factor of $\sim 1.6$.

 We suggest that the gas disk around TW Hya may be truncated compared to the dust disk
 to $\sim$ 120 AU, possibly due to photoevaporation.
  We  calculate the disk photoevaporative mass fluxes ($d\Sigma/dt$)
   due to FUV radiation from TW Hya.  TW Hya has a relatively high FUV luminosity $\sim 4 \times 10^{-3}$ L$_{\odot}$ (IUE Atlas, Valenti et al. 2003).
We use our disk temperature and density structure (Model I) and the analytical expressions of 
Adams et al. (2004)  to estimate the photoevaporation timescales for the TW Hya disk. We find that the photoevaporation timescales ($\Sigma/(d\Sigma/dt)$ where $\Sigma$ is the mass
surface density of the disk) due to FUV radiation decreases with increasing disk radius and that
the mass loss timescale at 120 AU is  $\sim 7 \times10^6$ years, similar to the age of the star/disk
system. 
 Therefore, it is plausible that during its lifetime, TW Hya has photoevaporated the disk
 outside of 120 AU,  carrying with it all gas and small ($<100 \mu$m) dust particles,
 but leaving behind the mm-sized dust grains seen in the sub-millimetre observations
 to extend to 174 AU. 
 
Table~\ref{twhyalin} lists other expected strong emission lines in our TW Hya disk model. 
We find that the [OI] line emission is very strong, well within the sensitivity limits of
the PACS instrument on  {\it Herschel}. PACS has a moderate resolving power ($\sim 1500$)
and hence will be unable to resolve the line, but is still capable of detecting the  line ($\sim
10^{-5}$ L$_{\odot}$)
above the dust continuum at these wavelengths ($\sim 10^{-4}$ L$_{\odot}$ at R$\sim 1500$.).

We end with a few caveats.  Our models are steady-state and do not account for viscous spreading of the disk and photoevaporative flows which may introduce time-dependent or advection effects
into the chemical and thermal structure. However, in the intermediate layer, the chemistry is
largely driven by rapid UV photodissociation and chemical timescales tend to be shorter than
dynamical timescales.  Also note that our solutions may not be unique, and use of a different dust model (that is still consistent with the disk SED) may yield different results. We also note here that we do not require depletion of CO as has been inferred by earlier modelers (van Zadelhoff et al. 2001, Qi et al. 2004, 2006).  We  expect it to be unlikely that the CO is frozen on cold dust grains, as the 
grains in the intermediate layer are sufficiently warm to thermally desorb CO ice mantles and the
UV field of TW Hya  is strong enough to keep the CO photodesorbed in the upper layers where the emission originates (also see Appendix B).

\section{Summary}
We present theoretical models of gas disks with optically thick dust around young stars.  Our emphasis is on the physical and chemical processes relevant in determining the vertical density and temperature structure of the gas accurately in the surface regions, where A$_{\rm V} \lesssim 10$ to the central star, and where
most infrared and sub-millimeter emission lines originate.  
We calculate expected luminosities of various diagnostic emission lines and
find that many infrared lines are detectable by PACS and HIFI on  {\it Herschel} and
by EXES on SOFIA. Some lines are also detectable by the IRS on the Spitzer Space Telescope.
Line luminosities are typically $\sim10^{-5}-10^{-6}$ 
L$_{\odot}$.  We find that decreased dust opacity in disks,  a result of grain growth 
and/or settling processes during evolution, can
increase [FeI]$24\mu$m line strengths, but that other lines are marginally affected.
X-rays heat the disk gas, and an absence of X-ray flux results in no [NeII]$12.8\mu$m
and [ArII]$7\mu$m emission  from the predominantly neutral disk surface below the EUV-ionized layer. X-rays also penetrate deeper into the disk and are important heating agents in the \av $\sim 1-3$ regions, and affect the  luminosities of the 
H$_2$ pure rotational lines, [FeI] and [SI] lines. The FUV luminosity of the central
star is  an important parameter, and in general,  heats the gas to increase line
emission.  [FeII] and [SiII], in particular, increase with higher FUV fields. 
 We find that the [OI]63$\mu$m line is a strong coolant and a good diagnostic
 probe of gas in disks.  The high line luminosity in all our disk models, $\sim
 10^{-4}$ L$_{\odot}$,  makes [OI] a good signature of the presence of  atomic
 gas and readily detectable by future facilities such
 as  {\it Herschel} and SOFIA. 
  
   We apply our models to the disk around TW Hya and can succesfully explain the observed CO and [NeII] emission from the disk. We suggest that the disk around TW Hya is being photoevaporated by its strong FUV radiation and that the gas disk is truncated ($\sim 120$ AU) relative to the dust disk ($\sim 174$ AU).

\section{Acknowledgements } 
We thank the referee for useful comments that have improved this paper.
We would like to thank Al Glassgold, Joan Najita and  Rowin Meijerink for many helpful 
discussions during the course of this
work, and for providing us with their results for model comparisons. 
We would also like to thank Charlie Qi for providing us with his CO data on TW Hya, 
and Michael Kaufman for many useful discussions.  We acknowledge financial support
by research grants  from  the NASA Origins of the Solar System Program (SSO04-0043-0032),
Astrophysics Theory Program (ATP04-0054-0083), and the NASA Astrobiology Institute. 
\appendix

\section{Numerical solutions of disk structure} 
Our numerical scheme is as follows. We divide the disk radial extent into logarithmic intervals in $r$, with approximately 200 annuli for each model.  We use the prescribed surface density power law to fix the surface density in each radial annulus.  As our models are 1+1D solutions, the disk solutions outward of a given radial position do not affect the disk structure interior to that radius. However,  column densities of gas and dust along the line of sight to the central star have to be calculated through the gas and dust in the inner radial zones, and we therefore proceed radially outward from the inner disk edge.  The first initial solution of the disk vertical structure is obtained by solving the dust radiative transfer and setting the gas and dust temperatures to be equal.   Starting from this initial guess for the density and temperature, we then solve for the chemistry and thermal balance in the disk and  compute the gas temperature as a function of vertical height in the disk, $z$.  
We begin our solution  at the midplane  and advance to the surface,  as this procedure ensures a more rapid convergence in the vertical density solution, which is constrained by the prescribed value for the surface density at that radius.

Many vertical  iterations are necessary for the density and temperature structure at a given radius in the disk to converge to a solution which gives the correct surface density and 
satisfies the condition for vertical hydrostatic equilibrium.   The condition for vertical hydrostatic equilibrium specifies the gas {\em pressure}, which may not yield a unique solution for the gas density and temperature at a spatial location $(r,z)$.  We find all density-temperature combinations that result in the required gas pressure at each grid cell, and that satisfy the physical condition of decreasing density with height.  For cells that yield multiple $n-T$ solutions, we arbitrarily choose the higher density solution. We also note that some disk models may have convectively unstable regions, where we set the temperature constant with $z$ until a stable solution is found at greater $z$.  The density in the convectively unstable region is set by the pressure criterion. 

The gas temperature is determined by simultaneously solving chemistry and imposing thermal balance.  Gas cooling is a strong function of the column densities of the different gas species to the surface of the disk. In calculating the escape probabilites, we assume two possible vertical directions perpendicular to the midplane of the disk as in some optically thin transitions the escape probability from a height $z$ through the disk midplane may be non-negligible.  We use the gas escaping column densities calculated from chemical abundances in the previous chemical solution vertically through the disk. The vertical grid spacing is calculated in an adaptive manner and is not held constant. This is particularly important in order to resolve important transitions such as the H/H$_2$ (which also changes the number density of gas)  and C/CO transitions as one approaches the disk surface.  We track important coolants such as CO and H$_2$O as abrupt gradients in their abundances can lead to sudden changes in gas temperature and erratic solutions. When the abundance of an important coolant begins to change with $z$, the grid spacing $\Delta z$ is made finer to resolve the transition and obtain a smooth solution. On an average $\sim 200$  vertical grid cells are  
required for each radial zone.  Vertical convergence is obtained with an average of 10-20 iterations and convergence is more rapid in the outer disk. Our final disk solution  is accurate to 2\% in temperature and surface density.

\section{Justification of Assumption of No Freeze-out in Intermediate Layer.}

There are two desorption mechanisms that help clear grains of ice mantles in the
intermediate layers of disks: (i) thermal desorption and (ii) FUV photodesorption.

\paragraph{Thermal desorption.} The grain temperature $T_{gr}$ at which ice mantles
thermally desorb can be calculated by equating the rate of gas phase molecules
hitting and sticking to a grain to the flux of thermally desorbing molecules from the
ice surface (see Hollenbach et al. 2008). 
\begin{equation}
n_s v_s \pi a^2 = \nu_s e^{-\Delta E_s/kT_{gr}} \left( { {4 \pi a^2} \over {A_s}}\right)
\end{equation}
Here, $n_s$ is the gas density of species $s$, $v_s$ is its thermal speed, $a$ is the
grain radius, $\nu_s$ is the vibrational frequency of the adsorbed molecule, 
$\Delta E_s$ is the binding energy of the adsorbed molecule, and $A_s$ is the area
occupied by a single adsorbed molecule. 
From this we find
\begin{equation}
T_{gr} = { {\Delta E_s/k} \over {\ln \left[ 4 \nu_s/(A_s n_s v_s)\right]}}
\label{tgr}
\end{equation}
Using $\Delta E_{H_2O}/k = 4800$K, $\Delta E_{CO}/k = 960$K, $\nu_{H_2O} = 
7.8 \times 10^{13} {\rm s}^{-1}, \nu_{CO} = 4.2 \times 10^{13} {\rm s}^{-1}$,
$A_{H_2O} \simeq A_{CO} \simeq10^{-15} {\rm \ cm}^2, n_{CO} \simeq
10^3 {\rm \ cm}^{-3}$, and $n_{H_2O} \simeq 1 {\rm \ cm}^{-3}$ (Hollenbach et al. 2008,
densities of H$_2$O and CO estimated from our models at $r=10$ AU and
\av=1), we find $T_{gr} (H_2O) \simeq 85$K and  $T_{gr} (CO) \simeq 20$K.
In our models, the largest (coldest) grains are 85 K at 10 AU and 20 K at $\sim$
140 AU. Therefore thermal desorption  prevents substantial freeze-out of CO
in the intermediate layers of nearly our entire fiducial disk, and prevents H$_2$O freeze-out
only for $r\lesssim 10$ AU. 

\paragraph{Photodesorption}
Hollenbach et al. (2008) discuss photodesorption of water ice in molecular clouds.
Photodesorption will clear an ice mantle when the photodesorption rate from a 
grain covered with ice equals the sticking rate of water molecules from the gas.
\begin{equation}
F_{FUV} Y \pi a^2 = n_{H_2O} v_{H_2O} \pi a^2
\end{equation}
where $F_{FUV}$ is the FUV photon flux and $Y \simeq 3 \times 10^{-3}$ is the
photodesorption yield from water ice (Westley et al. 1995, Andersson et al. 2006,
Hollenbach et al. 2008). $F_{FUV}$ is given approximately as
\begin{equation}
F_{FUV} = { {L_{FUV}} \over {h \nu_{FUV} 4 \pi r^2}} e^{-1.8 {\rm A}_{\rm V}}
\end{equation}
where $L_{FUV} = 10^{31} L_{31}$ erg s$^{-1}$ is the FUV luminosity of the central
star, and $h\nu_{FUV} \simeq 1.6 \times 10^{-11}$ ergs is the typical energy of
a photodesorbing FUV photon. From these equations, we find the critical radius
$r_{cr}$ inside of which water ice mantles are cleared due to phoodesorption
\begin{equation}
r_{cr} =\left( { {L_{FUV} Y } \over { 4 \pi h \nu_{FUV}  n_{H_2O} v_{H_2O} }}\right)^{1/2}
 e^{-0.9 {\rm A}_{\rm V}}
 \label{rcr}
\end{equation}
Note that a $z$ dependence enters because of the $z$ dependence of $n_{H_2O}$,
\av, and $v_{H_2O}$. Taking fiducial values of $n_{H_2O}=1$ cm$^{-3}$ and
$v_{H_2O} = 3 \times 10^4$ cm s$^{-1}$ we find
\begin{equation}
r_{cr} = 2000 L_{31}^{1/2} \left( {{n_{H_2O}} \over {1 {\rm \ cm}^{-3}}}\right)^{-1/2} {\rm AU \ at  \
 A}_{\rm V}=1 
\end{equation}
\begin{equation}
r_{cr} = 300 L_{31}^{1/2} \left( {{n_{H_2O}} \over {1 {\rm \ cm}^{-3}}}\right)^{-1/2} {\rm AU \ at  \
 A}_{\rm V}=3 
\end{equation}
Clearly, $r_{cr}$ depends on $n_{H_2O}$, which is a function of both $z$ and
$r$. However, it appears from the fiducial value of  $n_{H_2O}=1$ cm$^{-3}$ 
taken from our model at $r=10$ AU and at \av$=1$ that photodesorption should clear
our grains of water ice to a depth of \av $\lesssim 3$.

There is, however, an important caveat to these results. Water ice may form on
grain surfaces by the reaction of adsorbed O atoms with adsorbed H atoms, 
which forms adsorbed OH, followed by the reaction of adsorbed OH with
adsorbed H to form adsorbed H$_2$O (see Hollenbach et al. 2008). For this to
occur, the adsorbed O atom must not thermally desorb before an H atom strikes
a grain. This requirement translates to $T_{gr} \lesssim 36$K or $r\gtrsim 50$ AU for
\av $\sim 1$. Beyond 50 AU one can use Eq.~\ref{rcr} to estimate the critical
radius inside of which photodesorption clears ice mantles, but one needs to use
the larger of $n_{\rm O}$ or $n_{H_2O}$ instead of $n_{H_2O}$. Since  $n_{\rm O}
=10^3$cm$^{-3}$  at \av $=1$ and $r \sim 100$ AU, we find $r_{cr}\simeq 60 L_{31}^{1/2}$
AU. Therefore, there is the possibility of water ice freeze-out for \av $\gtrsim1$ and $r>r_{cr} \sim
60$AU. This could deplete the gas phase oxygen not in CO and reduce somewhat the predicted luminosities of those lines whose greatest contributions come from these regions. 
We have tested the sensitivity of our results to freeze-out by calculating the expected luminosity
of all our lines from our disk models where we omitted any emission from the 
freeze-out regions, implied by Equations~\ref{tgr} and \ref{rcr}. We find that the longer
wavelength transitions of water that arise mostly from low-lying states (at $E/k \lesssim 100$K)
can be less luminous than in Table~5 by factors ranging from $2-10$. The line
luminosities for these transitions listed in Table~5 should therefore be taken as upper limits,
and we have marked them appropriately.

\references
\reference{} Acke, B. \& van den Ancker, M.E. 2004, \aap, 426, 151 
\reference{} Adams, F.C., Hollenbach, D., Laughlin, G., Gorti, U. 2004, \apj, 611, 360
\reference{} Aikawa, Y., Umebayashi, T., Nakano, T., Miyama, S. 1997, \apjl, 486, L51
\reference{} Aikawa, Y., Umebayashi, T., Nakano, T., Miyama, S. 1999, \apj, 519, 705
\reference{} Aikawa, Y., Herbst E. 1999, \aap, 351, 253 
\reference{} Aikawa, Y., van Zadelhoff, G-J., van Dishoeck, E.  2002, \aap, 386, 622
\reference{} Aikawa, Y., Nomura, H. 2006, \apj, 642, 1152
\reference{} Alexander, R.D., Clarke, C.J., Pringle, J.E. 2005, \mnras, 358, 283
\reference{} Alexander, R.D., Clarke, C.J., Pringle, J.E. 2006, \mnras, 369, 216
\reference{} Andersson, S., 
Al-Halabi, A., Kroes, G.-J., \& van Dishoeck, E.~F.\ 2006, \jcp, 124, 4715
\reference{} Andrews, S. \& Williams, J.  2005, \apj, 631, 1134
\reference{} Andrews, S. \& Williams, J.  2007, \apj, 671, 1800
\reference{} Beckwith, S., Sargent, A., Chini, R., Guesten, R. 1990, \aj, 99, 924
\reference{}Bergin, E., Calvet, N., D'Alessio, P., Herczeg, G. 2003, \apjl, 591, L159
\reference{}Bouret, J.-C.,  \& Catala, C. 1998, \aap, 359, 1011 
\reference{}Calvet, N. \& Gullbring, E. 1998, \apj, 509, 802 
\reference{} Calvet, N., D'Alessio, P., Hartmann, L. et al. 2002, \apj, 568, 1008
\reference{}Chiang, E. \& Goldreich, P. 1997, \apj, 490, 368
\reference{} Chiang, E., Murray-Clay, R. 2007, Nature Physics, 3, 604
\reference{} Creech-Eakman, M.J., Chiang, E., Joung, R.M.K., et al. 2002, \aap, 385, 546
\reference{}D'Alessio, P., Canto, J., Calvet, N., Lizano, S. 1998, \apj, 500, 411
\reference{} Damjanov, I., Jayawardhana, R., Scholz, A. et al. 2007, \apj, 670, 1337
\reference{} Davis, S. 2007,\apj,660,1580
\reference{}Draine, B.  \& Li, A.  2001, \apj, 551, 807
\reference{}Dullemond, C.P., Dominik, C. \& Natta, A. 2001, \apj, 560, 957
\reference{} Dullemond, C.P., van Zadelhoff, G.J., Natta, A. 2002, \aap, 389, 464
\reference{}Dullemond, C.P. \& Dominik, C. 2005, \aap, 434, 971
Planets V, in press
\reference{}Dutrey, A., Guilloteau, S. \& Ho, P. 2006, in Protostars and Planets V, eds.  Reipurth, B.,
Jewitt, D., Keil, K., (Tucson: University of Arizona Press), p495
\reference{} Espalilat, C., Calvet, N., D'Alessio, P., Bergin, E. et al. 2007, \apjl, 664, L111 
\reference{}Feigelson, E. \&  Montmerle. T.  1999, \araa, 37, 363
\reference{}Feigelson, E., Getman, K., Townsley, L. et al. 2005, \apjs, 160, 401
\reference{}Furlan, E., Hartmann, L., Calvet, N. et al. 2006, \apjs, 165, 568
\reference{}Geers, V., Augereau, J.-C., Pontopiddan, K. et al. 2006, \aap, 459, 545  
\reference{}Gorti, U.  \& Hollenbach, D.  2004, \apj, 613, 424 (GH04)
\reference{} Glasse, A.C., Atad-Ettedgui, E.I., Harris, J.W. 1997, SPIE, 2871, 1197 
\reference{}Glassgold, A., Najita, J. \& Igea, J. 1997, \apj, 485, 344 
\reference{}Glassgold, A., Najita, J. \& Igea, J. 2004, \apj, 615, 972
\reference{}Glassgold, A., Najita, J. \& Igea, J. 2007, \apj,  656, 515  
\reference{} G\"{u}del, M., Briggs, K.R., Arzner, K. et al. 2007, \aap, 468, 353
\reference{}Gullbring, E., Hartmann, L., Briceno, C., Calvet, N. 1998, \apj, 492, 323
\reference{} Gullbring, E., Calvet N., Muzerolle, J., Hartmann, L. 2000, \apj, 544, 927
\reference{}Habart, E., Natta, A. \& Krugel, E. 2004, \aap, 427, 179
\reference{}Haisch, K. et al. 2001, \aj, 121, 1512
\reference{}Hartmann, L., Calvet, N., Gullbring, E., D'Alessio, P. 1998, \apj, 495, 385
\reference{}Herzceg, G., Wood, B., Linsky, J. et al. 2004, \apj, 607, 369
\reference{} Hillenbrand, L.A. 2007, to appear in ``A Decade of Discovery: Planets Around Other Stars" STScI Symposium Series 19, ed. M. Livio
\reference{} Hollenbach, D., Takahashi, T., Tielens, A.G.G.M. 1991, \apj, 377, 192
\reference{}Hollenbach, D., Johnstone, D., Lizano, S., Shu, F. 1994, \apj, 428, 654
\reference{} Hollenbach, D., Tielens, A.G.G.M., 1999, Rev. of Mod. Phys., 71, 173
\reference{}Hollenbach, D., Gorti, U., Meyer, M. et al. 2005, \apj, 631, 1180
\reference{} Hollenbach, D., Kaufman, M., Bergin, E., Melnick, G. 2008, in preparation
\reference{} Hughes, A.M., Wilner, D.J., Calvet, N. et al. 2007, \apj, 664, 536
\reference{}Jonkheid, B., Faas, F. G. A., van Zadelhoff, G.-J., van Dishoeck, E.F. 2004, \aap, 428, 511
\reference{}Kamp, I. \& Dullemond C.P. 2004, \apj, 615, 991
\reference{} Kastner, J., Zuckerman, B.,  Weintraub, D., Forveille, T. 1997, Science, 277, 67
\reference{} Kaufman, M.J., Wolfire, M.G., Hollenbach, D.J., Luhman, M.L. 1999, \apj, 527, 795
\reference{}Kessler-Silacci, J., Augereau, A., Dullemond, C.P. et al. 2006, \apj, 639, 275
\reference{} Lacy, J., Richter, M., Greathouse, T., et al.  2002, PASP, 114, 153
\reference{} Lahuis, F., van Dishoeck, E., Blake, G., et al. 2007, ApJ, 665, 492 
\reference{}Li, A.  \& Draine, B.  2001, \apj, 554, 778
\reference{} Luhman, K., Adame, L., D'Alessio, P. et al. 2007, ApJL, 666, 1219 
\reference{} Maloney, P., Hollenbach, D., Tielens A.G.G.M 1996, \apj, 466, 561
\reference{} Mamajek., E. 2005, \apj, 634, 1385
\reference{} Markwick, A., Ilgner, M., Millar, T.J., Henning, Th. 2002, \aap, 385, 632  
\reference{} Meijerink, R., Glassgold, A., Najita, J. 2007, \apj, in press (astroph: 0712.0112)
\reference{} Muzerolle, J., Calvet, N., Briceno, C. et al. 2000, \apjl, 535, L47
\reference{} Muzerolle, J., Hillenbrand, L., Calvet, N. et al. 2003, \apj, 592, 266
\reference{}Muzerolle, J., Adame, L., D'Alessio, P.  2006, \apj, 643, 1003
\reference{} Najita, J., Strom, S.E., Muzerolle, J. 2007, MNRAS, 378, 369 
\reference{} Natta, A., Prusti, T., Neri, R., et al. 2001, \aap, 371, 186
\reference{} Nomura, H., Millar, T.J. 2005, \aap, 438, 923
\reference{} Nomura, H., Aikawa, Y., Tsujimoto, M., Nakagawa, Y., Millar, T., 2007, \apj, 661, 334
\reference{}Pascucci, I., Gorti, U., Hollenbach, D. et al. 2006, \apj, 651, 1177
\reference{} Pascucci, I., Hollenbach, D., Najita, J. et al. 2007, ApJ, 663, 383 
\reference{}Preibisch, T., Kim, Y.-C., Favata, F. et al. 2005, \apjs, 160, 401
\reference{} Qi, C., Wilner, D.J., Calvet, N. et al. 2006, \apjl, 636, L157
\reference{} Qi, C., Ho, P.T.P., Wilner, D.J. et al. 2004, \apjl, 616, L11
\reference{} Quillen, A., Blackman, E., Frank, A., Varniere, P. 2004, ApJL, 612, 137
\reference{}Rafikov, R. \&  De Colle, F.  2006, \apj, 646, 275
\reference{} Savage, B.D., Sembach, K.R. 1996, \apj, 470, 893
\reference{}Schreyer, K., Semenov, D., Henning, Th., Fobrich J.  2006, \apjl, 637, L129 
\reference{} Semenov, D., Wiebe, D., Henning, Th. 2006, \apjl, 647, L57
\reference{} Shuping, R., Kassis, M., Morris, M et al. 2006, \apjl,  644, L71
\reference{} Sicilia-Aguilar, A., Hartmann, L., Calvet, N. et al. 2006, \apj, 638, 897
\reference{}Siess, L., Dufour, E. \& Forestini, M. 2000, \aap, 358, 593
\reference{}Steltzer, B., Flaccomio, E., Montmerle, T. et al. 2005, \apjs, 160, 557
\reference{}Telleschi, A., Guedel, M., Briggs, K.R., Audard, M., Scelsi, L. 2007, \aap, 468, 541
\reference{} Uchida, T., Calvet, N., Hartmann, L. et al. 2004, \apjs, 154, 439
\reference{} Udry, S., Fischer, D., Queloz, D.  2007, in Protostars and Planets V, eds.  Reipurth, B.,
Jewitt, D., Keil, K., (Tucson: University of Arizona Press), p685
\reference{}Valenti, J.A., Fallon, A. \& Johns-Krull, C. 2003, \apjs, 147, 305
\reference{}van Boekel, R., Min, M., Waters, L.B.F.M. et al. 2005, \aap, 437, 189
\reference{} van Boekel, R., Waters, L.B.F.M., Dominik, C. et al. 2004, \aap, 418, 177  
\reference{} van Zadelhoff, G.-J., van Dishoeck, E., Thi, W-F., Blake, G.A. 2001, \aap, 377, 566
\reference{}Watson, A., Stapefeldt, K., Wood, K., Menard, F. 2007, in Protostars and Planets V, 
eds.  Reipurth, B.,
Jewitt, D., Keil, K., (Tucson: University of Arizona Press), p523 
\reference{} Watson, D.,  Bohac, C.J., Hull, C. et al. 2007, Nature, 448, 1026
\reference{}Weingartner, J.  \&  Draine, B.  2001, \apjs, 134, 263
\reference{} Westley, M., Baragiola, R., Johnson, R., Baratta, G. 1995, Nature, 373, 405
\reference{} Willacy, K., Klahr, H., Millar, T.J., Henning, Th. 1998, \aap, 338, 995
\reference{} Willacy, K., Langer, W.D.  2000, \apj, 544, 903
\reference{} Willacy, K. 2007, \apj, 660, 441
\reference{}Wilms, J., Allen, A. \& McCray, R.  2000, \apj, 542, 914
\reference{} Webb, R. A., Zuckerman, B., Platais, I., et al. 1999, ApJ, 512, 63 
\reference{}Zuckerman, B., Forveiile, T., Kastner, J. 1995, Nature, 373, 494
\newpage

\begin{table}[t]
\caption{Stellar Input Parameters For Fiducial 1M$_{\odot}$ Case}
\center
\begin{tabular}{ll}
\tableline
\tableline
Mass & 1 M$_{\odot}$ \\
Radius &  2.61 R$_{\odot}$ \\
Temperature & 4278 K \\
Luminosity & 2.34 L$_{\odot}$ \\
Accretion rate & $3.0 \times 10^{-8}$ M$_{\odot}/$yr \\
Log L$_{\rm FUV}$ (erg s$^{-1}$) & 31.7 \\
Log L$_{\rm X}$ (erg s$^{-1}$) & 30.4 \\
Log $\phi_{\rm EUV}$  (s$^{-1}$) & 41.6 \\
\tableline
\end{tabular}
\label{starpar}
\end{table}

\begin{table}[f]
\centering
\caption{Gas Phase Elemental Abundances}
\label{abun}
\begin{tabular}{lc}
&\\
\tableline
\tableline
Element & Gas Phase Abundance\\
\tableline
H &  1.0 \\
 He & $0.1$ \\
 C & $1.4 \times 10^{-4}$ \\
 O & $3.2 \times 10^{-4}$ \\
 Ne & $1.2 \times 10^{-4}$ \\
 Mg & $1.1 \times 10^{-6}$ \\
 S & $2.8 \times 10^{-5}$ \\
 Si & $1.7 \times 10^{-6}$ \\
 Fe & $1.7 \times 10^{-7}$ \\
 Ar & $ 6.3 \times 10^{-6}$ \\
\tableline
\end{tabular}
\end{table}

\begin{table}[t]
\centering
\caption{Fiducial Disk Model - Input Parameters}
\label{fidpar}
\begin{tabular}{ll}
\tableline
\tableline
Disk mass & 0.03 M$_*$ \\
Surface density & $\Sigma(r) \propto r^{-1}$  \\
Inner disk radius & 0.5 AU \\
Outer disk radius & 200 AU \\
Gas/Dust Mass Ratio & 100 \\
Dust size distribution & $n(a) \propto a^{-3.5} $  \\
Min. grain size $a_{min}$ & $ 50 $\AA   \\
Max. grain size $a_{max}$ & $20 \mu $m  \\
$\sigma_{\rm H}$ & $2 \times 10^{-22} {\rm cm}^2/{\rm H}$ \\
PAH abundance/H  & $8.4 \times 10^{-8}$  \\
\tableline
\end{tabular}
\end{table}

\begin{table}
\centering
\caption{Predicted Line Luminosities  (Log L/L$_{\odot}$) }
\label{fidline}
\begin{tabular}{lcccc}
\tableline
\tableline
Line &  Model A & Model B & Model C & Model D \\
\hline
 & (Fiducial) & $ 0.01 \times \sigma_{\rm H }$
 & L$_{\rm X}=0 $ & $10 \times$  L$_{\rm FUV} $ \\
\tableline
ArII 7$\mu$m &  -5.87& -6.25 &  --- & -5.50\\
H$_2$ S(2) 12$\mu$m &  -4.82 & -4.97 & -5.10&-4.40 \\
NeII 12.8$\mu$m          &  -5.54 & -5.83 &  --- & -5.28\\
H$_2$ S(1) 17$\mu$m &  -4.53& -4.48& -4.85& -4.22\\
FeI 24$\mu$m              &  -6.03& -4.62 & -.6.41& -5.94\\
SI 25$\mu$m             &  -5.39 & -5.55 & -5.84& -4.9\\
FeII  26$\mu$m            &  -6.47 & -5.96 & -6.42& -5.55\\
H$_2$ S(0) 28$\mu$m &  -5.14& -5.16 & -5.31& -4.79\\
H$_2$O (total cooling)   & -4.46    &  -4.08& -4.55 & -4.26\\
SiII 35$\mu$m           &  -5.87& -5.84& -6.01& -4.94\\
OI 63$\mu$m          &  -4.34& -4.38 & -4.53 & -3.62\\
CII 158$\mu$m          &  -6.30& -6.04& -6.38 & -5.44 \\
CI 371$\mu$m          &  -6.60& -6.42& -7.14 & -6.27\\
CO 2-1           &  -6.98& -7.19& -6.96 & -6.89\\
CO 3-2           &  -6.55& -6.79& -6.42 & -6.33 \\
CO 6-5           &  -5.92& -5.84& -5.96 & -5.72 \\
\tableline
\end{tabular}
\end{table}

%

 \begin{table}
 \caption{H$_2$O Emission Line Luminosities For  Model Disk A 
 \label{wlines}} 
 \begin{minipage}{3in}
\begin{tabular}{c r l}
\tableline
\tableline
Transition\tablenotemark{a} & $\lambda(\mu$m) & Log (L/L$_{\odot}$) \\
\tableline
p $ 5_{ 3 3}- 4_{ 0 4}$&      35.471 &       -6.365  \\
p $ 6_{ 5 1}- 5_{ 4 2}$&      35.904 &       -6.379   \\
o $ 6_{ 5 2}- 5_{ 4 1}$&      35.937 &       -6.386   \\
p $ 6_{ 2 4}- 5_{ 1 5}$&      36.212 &       -6.381   \\
o $ 5_{ 5 0}- 4_{ 4 1}$&      39.374 &       -6.128   \\
p $ 5_{ 5 1}- 4_{ 4 0}$&      39.379 &       -6.128   \\
p $ 6_{ 4 2}- 5_{ 3 3}$&      39.398 &       -6.234   \\
o $ 6_{ 4 3}- 5_{ 3 2}$&      40.336 &       -6.259   \\
o $ 4_{ 3 2}- 3_{ 0 3}$&      40.690 &       -6.143 $\downarrow$   \\
p $ 6_{ 3 3}- 5_{ 2 4}$&      40.759 &       -6.205   \\
o $ 5_{ 4 1}- 4_{ 3 2}$&      43.893 &       -6.002   \\
p $ 5_{ 4 2}- 4_{ 3 1}$&      44.194 &       -6.012   \\
o $ 5_{ 2 3}- 4_{ 1 4}$&      45.111 &       -5.904  \\
p $ 3_{3 1}-2_{0 0}$ &        46.483  &     -6.354  $\downarrow$ \\
o $ 5_{ 3 2}- 4_{ 2 3}$&      47.972 &       -5.923   \\
p $ 4_{ 4 0}- 3_{ 3 1}$&      49.281 &       -5.805   \\
o $ 4_{ 4 1}- 3_{ 3 0}$&      49.336 &       -5.795  $\downarrow$ \\
o $ 6_{ 3 4}- 5_{ 2 3}$&      49.389 &       -6.271   \\
p $ 5_{ 3 3}- 4_{ 2 2}$&      53.137 &       -5.991    \\
p $ 4_{ 3 1}- 3_{ 2 2}$&      56.324 &       -5.728   $\downarrow$ \\
p $ 4_{ 2 2}- 3_{ 1 3}$&      57.636 &       -5.754   $\downarrow$ \\
o $ 4_{ 3 2}- 3_{ 2 1}$&      58.698 &       -5.772   $\downarrow$ \\
o $ 8_{ 1 8}- 7_{ 0 7}$&      63.322 &       -6.474   \\
p $ 8_{ 0 8}- 7_{ 1 7}$&      63.456 &       -6.475   \\
o $ 6_{ 2 5}- 5_{ 1 4}$&      65.164 &       -6.254   \\
o $ 3_{ 3 0}- 2_{ 2 1}$&      66.436 &       -5.685 $\downarrow$ \\
p $ 3_{ 3 1}- 2_{ 2 0}$&      67.088 &       -5.783  $\downarrow$ \\
p $ 5_{ 2 4}- 4_{ 1 3}$&      71.066 &       -5.991 $\downarrow$ \\
 \tableline
\end{tabular} 
\end{minipage}
\begin{minipage}{3in}
\begin{tabular}{c r l}
\tableline
\tableline
Transition & $\lambda(\mu$m) & Log (L/L$_{\odot}$) \\
\tableline
p $ 7_{ 1 7}- 6_{ 0 6}$&      71.538 &       -6.281   \\
o $ 7_{ 0 7}- 6_{ 1 6}$&      71.946 &       -6.283   \\
o $ 3_{ 2 1}- 2_{ 1 2}$&      75.379 &       -5.565   $\downarrow$ \\
o $ 4_{ 2 3}- 3_{ 1 2}$&      78.740 &       -5.874   $\downarrow$ \\
o $ 6_{ 1 6}- 5_{ 0 5}$&      82.030 &       -6.052   \\
p $ 6_{ 0 6}- 5_{ 1 5}$&      83.282 &       -6.088   \\
p $ 3_{ 2 2}- 2_{ 1 1}$&      89.987 &       -5.778   $\downarrow$ \\
p $ 5_{ 1 5}- 4_{ 0 4}$&      95.625 &       -5.860   $\downarrow$ \\
o $ 5_{ 0 5}- 4_{ 1 4}$&      99.491 &       -5.849   $\downarrow$ \\
o $ 5_{ 1 4}- 4_{ 2 3}$&     100.911 &       -6.448    \\
p $ 2_{ 2 0}- 1_{ 1 1}$&     100.980 &       -5.765   $\downarrow$ \\ 
o $ 2_{ 2 1}- 1_{ 1 0}$&     108.071 &       -5.758  $\downarrow$ \\
o $ 4_{ 1 4}- 3_{ 0 3}$&     113.536 &       -5.916 $\downarrow$ \\
p $ 4_{ 0 4}- 3_{ 1 3}$&     125.353 &       -5.912   $\downarrow$ \\
p $ 3_{ 3 1}- 3_{ 2 2}$&     126.711 &       -6.401  $\downarrow$ \\
o $ 4_{ 2 3}- 4_{ 1 4}$&     132.405 &       -6.165   $\downarrow$ \\
p $ 3_{ 3 0}- 3_{ 2 2}$&     136.492 &       -6.454   $\downarrow$ \\
p $ 3_{ 1 3}- 2_{ 0 2}$&     138.524 &       -5.728   $\downarrow$ \\
p $ 3_{ 2 2}- 3_{ 1 3}$&     156.193 &       -6.150  $\downarrow$ \\
o $ 3_{ 0 3}- 2_{ 1 2}$&     174.620 &       -5.947   $\downarrow$ \\
o $ 2_{ 1 2}- 1_{ 0 1}$&     179.523 &       -6.070   $\downarrow$ \\
o $ 2_{ 2 1}- 2_{ 1 2}$&     180.482 &       -5.934   $\downarrow$ \\
p $ 2_{ 2 0}- 2_{ 1 1}$&     243.969 &       -6.223    $\downarrow$ \\
o $ 3_{ 1 2}- 2_{ 2 1}$&     259.985 &       -6.370   $\downarrow$ \\
p $ 1_{ 1 1}- 0_{ 0 0}$&     269.268 &       -6.378   $\downarrow$ \\
o $ 3_{ 1 2}- 3_{ 0 3}$ &   273.196 &       -5.336 \\
p $ 2_{ 0 2}- 1_{ 1 1}$&     303.447 &       -6.372    $\downarrow$ \\
p $ 2_{ 1 1 }- 1_{ 0 0 }$&     398.636 &       -6.499   $\downarrow$ \\
 \tableline
\end{tabular}
\end{minipage}
\tablenotetext{a}{ The symbols 'o' and 'p' denote ortho-H$_2$O and  para-H$_2$O transitions respectively.}
\tablenotetext{} {$\downarrow $These lines may be affected by water ice formation on
grains and are upper limits}
\end{table}

\begin{table}
\centering
\caption{Model Parameters For Disk Around TW Hya}
\label{twhyapar}
\begin{tabular}{ll}
\tableline
\tableline
Spectral Type & K7 \\
Mass & 0.77 M$_{\odot}$ \\
Stellar Temperature & 4000 K \\
Stellar Radius & 1 R$_{\odot}$  (as in  Calvet et al. 2002) \\ 
X-ray Luminosity & $ 2.3 \times 10^{30}$ erg s$^{-1}$ \\
UV Spectrum & IUE data from Valenti et al. 2003 \\
& \\
Dust Model & From Calvet et al. 2002 \\
$a_{min}$ & 0.01 $\mu$m \\
$a_{max}$ & 1 cm \\
Disk Inner Radius & 4 AU \\
Disk Outer Radius & 174 AU \\
Dust/Gas Mass Ratio & 0.01 \\
Total Disk Mass & 0.03 M$_{\odot}$ \\
\tableline
\end{tabular}
\end{table} 

\begin{table}
\centering
\caption{Comparison of Model Line Luminosities (in L$_{\odot}$)
 and Observational Data.
\label{twhyalin}}
\begin{tabular}{l c c c }
\tableline
\tableline
& Model I   & Model II  & Obs. Data\\
&  (R$_{out}$=174 AU) & (R$_{out}$=120AU) &\\
\tableline
CO 2-1 &  $1.8 \times 10^{-8} $    &  $8.4 \times 10^{-9} $    &  
\tablenotemark{a}$7.9 \times 10^{-9} $   \\  
CO 3-2 &  $5.1 \times 10^{-8} $    &  $2.4 \times 10^{-8} $    & 
\tablenotemark{a} $2.4 \times 10^{-8} $   \\  
CO 6-5 &  $7.8 \times 10^{-8} $    &  $7.5 \times 10^{-8}$   & 
\tablenotemark{a} $4.6 \times 10^{-8} $   \\  
NeII 12.8$\mu$m &  $9.1 \times 10^{-6} $    &  $9.1 \times 10^{-6} $   &  
\tablenotemark{b}$8.6 \times 10^{-6} $   \\  
\tableline
H$_2$ S(1) 17$\mu$m &  $2.0 \times 10^{-7} $  &  $2.0 \times 10^{-7} $   \\
SI 25.2$\mu$m        &  $3.0 \times 10^{-6} $       &  $3.0 \times 10^{-6} $   \\ 
OI 63$\mu$m        &  $1.3 \times 10^{-5} $    &  $1.0 \times 10^{-5} $   \\ 
OI 145$\mu$m      &  $3.3 \times 10^{-6} $      &  $2.0 \times 10^{-6} $   \\ 
CII 158$\mu$m   &  $9.9 \times 10^{-7} $         &  $6.7 \times 10^{-7} $   \\ 
CI 371$\mu$m        &  $1.1 \times 10^{-7} $    &  $5.9 \times 10^{-8} $   \\ 
\tableline
\end{tabular}
\tablenotetext{a}{Qi et al. 2006}
\tablenotetext{b}{Uchida et al. 2004}
\end{table}
\newpage

\end{document}